\documentclass{aa} 
\usepackage{graphicx}
\usepackage{txfonts}
\usepackage{hyperref} 
\hypersetup{
 colorlinks=true,
 linkcolor=blue,
 urlcolor=cyan,
 citecolor=magenta,
 pdftitle={Sharelatex Example},
}
\usepackage{subfig}
\usepackage{natbib}
\bibpunct{(}{)}{;}{a}{}{,}
\begin{document}

 \title{Navarro-Frenk-White dark matter profile and the dark halos around disk systems}

 \author{R. Dehghani
 \inst{1,2}
 P. Salucci\inst{1}
 \and
 H. Ghaffarnejad \inst{2}
 }

 \institute{SISSA, Via Bonomea, 265, 34136 Trieste, Italy\\
 \email{salucci@sissa.it}
 \and
 Faculty of Physics, Semnan University, Semnan, 35131-19111, Iran\\
 \email{Razieh-Dehghani@semnan.ac.ir}
, hghafarnejad@semnan.ac.ir }

 \date{Received mm dd, yy; accepted mm dd, yy}
 
 \abstract
 {The $\Lambda$ cold dark matter ($\Lambda$CDM) scenario is able to describe the Universe at large scales, but clearly shows some serious difficulties at small scales. The core-cusp  question is one of these difficulties: the inner dark matter (DM) density profiles of spiral galaxies generally appear to be cored, without the $r^{-1}$ profile that is predicted by N-body simulations in the above scenario.}
 {It is well known that in a more physical context, the baryons in the galaxy might backreact and erase the original cusp through supernova explosions. Before the efficiency and the presence of this effect is investigated, it is important to determine how wide and frequent the discrepancy between observed and N-body-predicted profiles is and what its features are.} 
 {We used more than 3200 quite extended
rotation curves (RCs) of good quality and high resolution of disk systems that included normal and dwarf spirals as well as low surface brightness galaxies. The curves cover all magnitude ranges. All these RCs were condensed into 26 coadded RCs, each of them built with individual RCs of galaxies of similar luminosity and morphology. We performed mass models of these 26 RCs using the Navarro-Frenk-White (NFW) profile for the contribution of the DM halo to the circular velocity and the exponential Freeman disk for the contribution of the stellar disk.}
 {The fits are generally poor in all the 26 cases: in several cases, we find $\chi^2_{red}>2$. Moreover, the best-fitting values of three parameters of the model ($c$, $M_D$, and $M_{vir}$) combined with those of their 1$\sigma$ uncertainty clearly contradict well-known expectations of the $\Lambda$CDM scenario. We also tested the scaling relations that exist in spirals with the outcome of the current mass modeling: the modeling does not account for these scaling relations.} 
 {The results of testing the NFW profile in disk systems indicate that this DM halo density law cannot account for the kinematics of the whole family of disk galaxies. It is therefore mandatory for the success of the  $\Lambda CDM$ scenario in any disk galaxy of any luminosity or maximum rotational velocity to transform initial cusps into the observed cores. }


 \maketitle
%

\section{Introduction}
Dark matter (DM), which makes up 85\%\ of all the matter in the Universe, is still one of the most elusive mysteries in present-day physics. The existence of nonbaryonic DM is considered a fact that has been proved by  Planck measurements of the cosmic microwave background (CMB), for instance. This component is thought to be made of nonrelativistic particles that can be described by a collisionless fluid that interacts with ordinary particles (baryons) through gravity alone because the cross section for nucleons is thought to be only $\sim 10^{-26} cm^2$ \citep [e.g.,][]{Jungman.Kam.16}.

At present, $\Lambda$ cold dark matter ($\Lambda$CDM) is known as a very successful cosmological  scenario in describing the formation and evolution of the large-scale structures of the Universe \citep{Frenk.Whi.85, Bode.Ost.01}. The N-body simulations in this scenario have been able to resolve the current structure of virialized objects from clusters to dwarf galaxies. 
 The simulations have shown that the $\Lambda$CDM power spectrum of the density perturbations combined with the collisionless nature of the CDM particles lead in any virialized object from clusters to dwarf galaxies to a very peculiar cuspy dark halo density profile : $\rho_{DM}\propto r^{-1}$. In detail, dark halos around galaxies are  well described by the Navarro-Frenk-White (NFW) profile \citep{NFW96, NFW2.96}. 
 
 It is also known that this sharp central cusp of the DM density distribution strong contradicts the analysis of the kinematics of disk galaxies. For the first time, \citet{Gentile.Sal.04} showed in mass models obtained from six high-quality extended RCs that combined optical and $2\text{D}$ HI data overwhelming evidence for cored dark halo distributions beyond any statistical or bias uncertainty.  More references on this topic is given in \citet[]{Salucci.19}.

 About 100 RCs of the above quality have been thoroughly investigated so far, and the results have led to a similar decisive support in favor of the DM halo cored distribution \citep[e.g.,][]{Adams.12, Donato.Gen.09, Weinberg.15, Simon.Bol.05, Deblok.Bos.02, Spekkens.05}. The current status of the analysis of individual RCs is described in \citet{Korsega.18, Korsega.19}, who  obtained mass models of 31 spiral and 
 irregular galaxies,  derived using hybrid rotation curves that combined high-resolution optical GHASP Fabry-Perot $H \alpha$ RCs with extended radio WHISP $HI$ RCs. Moreover, for each galaxy, the analysis took advantage of high-quality 3.4 $\mu m$ WISE photometry, which is a fair indicator of the stellar disk mass. Korsaga and collaborators found that independent of the value of the galaxy optical velocity ($V_{opt}$), the performance of the baryonic matter plus NFW halo in reproducing the RCs was poor. Cored DM halo models are also required in galaxies that are dominated by luminous matter inside $R_{opt}$. This topic has been considered serious enough for alternatives to the collisionless dark particle scenario to have been proposed, including warm dark matter (WDM), e.g.     \cite{Dipaolo.Nest.18}, self-interacting DM \citep{Spergel.Ste.00}, and ultra-light axions (ULAs) \citep[e.g.,]{Tulin.13}; moreover, the possibility that the dark particles gain energy from the standard model particles by means of feedbacks, that is, by sudden gas outflows into the galaxy halo driven by supernova explosions, raises doubts on the collisionless status of dark particles in the $\Lambda$CDM scenario \citep[e.g.,]{Governato.12}. 
\
Although the cored density distribution is often considered as preponderant in disk systems, a complete and comprehensive comparison between the galaxy kinematical data and the NFW halo profile has not been carried out so far. The notable questions even in disk systems are  \textbf{i)} the degree of the disagreement of data with{\it } simulations as a function of the galaxy reference velocity ($V_{opt}$) and of its Hubble type, and \textbf{ii)} the way in which this discrepancy  can be qualitatively characterized in general and in the various different objects. 
\
The results found in this work are a necessary first step to answer these questions. In this regard, we stress that we tested the NFW profile not only by determining whether it reproduces the kinematics of disk systems, but also by verifying whether the resulting best-fit models {\bf i)} reproduce the relationship between two profile parameters, that is, concentration ($c$) and virial mass ($M_{vir}$), that emerges from N-body simulations; and if they {\bf ii)} lead to fractions of dark matter in systems of different virial masses that agree with independent measurements.
\
 Only a complete assessment of the actual inability of the virialized noncollisional matter to describe the actual density halos around galaxies can shed light on how the collisionless scenario may be changed properly. The processes within a collisionless particle scenario include adiabatic contraction, feedback from supernovae, and clump migration
\citep[see, e.g.,][]{Blumenthal.Fab.86,Elzant.Fre.16,Macci.Sti.12, Dicintio.14}. Moreover, newly proposed scenarios for the DM particles do not always produce cores. For instance, the ULAs scenario seems to form core radii in the DM halos with $M<10^{11} M_\odot $ \citep[e.g.,][]{deMartino.Bro.18}; in the baryonic feedback scenario, the DM halos retain their cuspy profiles if $M_{stellar}/M_{halo}<0.001$ or $M_{stellar}/M_{halo}> 0.005;$ and in WDM scenario, it is difficult to explain cores in the halo densities of dwarf galaxies \citep[e.g.,][]{Salucci.19}.
\
Furthermore, we also exploit the fact that the luminous and dark masses of galaxies can be determined by methods alternative to the RC analysis and then compared with those obtained from the RC analysis. In detail, disk masses can be derived by multiplying the galaxy luminosities with their mass-to-light ratios estimated from their colors, while the  virial masses are obtained by means of weak-lensing measurements \citep[e.g.,][]{Reyes.12,Mandelbaum.16} or from the abundance-matching method \citep[e.g.,][]{Moster.Na.13,Shankar.Lap.06,Rodriguez.12}. 
We assume $\Omega_{total}=1$; $\Omega_{matter}=0.3$; $\Omega_{baryonic}=0.04$; $\Omega_{DM}=0.26$; $\Lambda=0.7$ and  $H_0= 72 \ km\ s^{-1} Mpc^{-1}$.

 The structure of this paper is as follows: in  Section \ref{sam},  Sect. \ref{stack}, we describe the samples we used. In Sect. \ref{coadd} we review the process of coaddition of individual RCs, and in Sect. \ref{mmod} we discuss the mass modeling, which includes the disk and halo components. In Section \ref{result} we obtain the mass model of 26 coadded RCs and discuss their implications, in particular, the implications from the resulting best-fitting values of the model-free parameters, $M_D,M_{vir}$ , and $c$. Finally, we draw conclusions in Section \ref{conclu} based on our results.

 \begin{figure*}[ht!]
\begin{center}\label{fig11}
\includegraphics[scale=0.9]{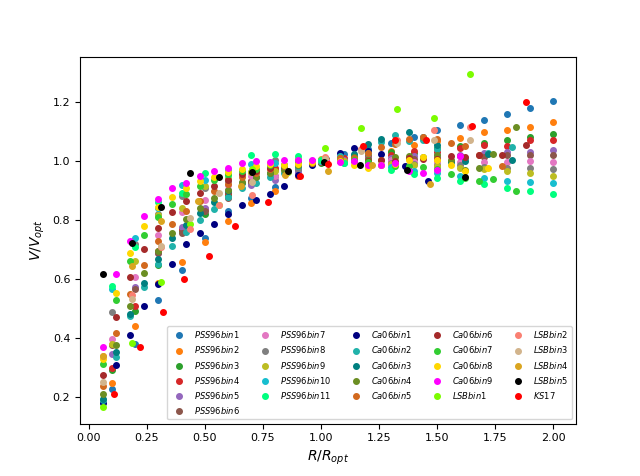}
\end{center}
\caption{Twenty-six double-normalized coadded RCs. The data and their uncertainties  are listed in the online table.}
\label{obdat}
\end{figure*}

\section{Samples and mass model}\label{sam}

\subsection{Sample}\label{stack}
The 26 coadded RCs that we obtained from more than 3200 individual RCs of disk systems and that we discuss in Sect. \ref{coadd} belong to four samples: \citet{PSS96}, \citet{ Catinella.Gio.06}, \citet{Dipaolo.Sal.18}, and \citet{Karukes.Sal.17}. These coadded RCs have never been used for the goal of testing NFW profile. 900 RCs in \citet{PSS96} were also independently analyzed by \citet{Catinella.Gio.06}. \textbf{i)} In the first sample, optical RCs out to $\sim 1.2\ R_{opt}$ are combined with HI RCs out to $2\, R_{opt}$ , where $R_{opt}$ is the optical radius and defined as the radius enclosing $83\%$ of the  total stellar mass of a galaxy. \textbf{ii)} The second sample is composed of optical RCs alone. \textbf{iii)} The third and the fourth samples consist of a combination of about $90\%$ optical RCs and 10\% HI RCs. Then, the innermost region ($R<2/3\ R_{opt}$) of almost all the 3200  galaxies are traced by high-resolution optical RCs.
In short, each coadded RC is obtained from a large number of measurements, all of which belong to RCs of galaxies of same Hubble type and with similar values of $V_{opt}$ , which is the velocity at the optical radius ($R_{opt}$). Moreover, they have a very high spatial resolution, so that they are not affected by the beam-smearing effects that may distort   the inner profile of low-resolution RCs. Finally, all the coadded RCs we used are extended out to at least, $R\simeq 3\ R_D$ , where $R_D$ is the disk scale length. This is discussed in Sect. \ref{coadd}.

The first sample has been published in \citet{PSS96} (hereafter \textbf{PSS96}). It consists of $\sim 900$ individual RCs, arranged into 11 M$_I$ (the I-band absolute magnitude ) bins, whose central values span from -17.5 to -23 (which corresponds to $75~ km\ s^{-1}$ to $279~ km\ s^{-1}$), and which yield 11 coadded rotation curves built by arranging the RCs in their corresponding  $M_I$ bins. In radial normalized coordinates $x\equiv R/R_{opt}$, the radial bins have a size of $\sim 0.1$.

 The second sample comes from \citet{Catinella.Gio.06} \citep[hereafter \textbf{Ca06}; see also][]{Lapi.Sal.18}, and it consists of $\sim 2200$ individual RCs that also include the $\bf{PSS96}$ sample. Individual RCs are arranged into nine optical velocity bins, whose central values span from $104~km\ s^{-1}$ to $330~km\ s^{-1}$ , and the normalized radial bin has a size of $0.06$. 

The third sample, from \citet{Dipaolo.Sal.18} (hereafter \textbf{LSB}), consists of $72$ low surface brightness galaxies with velocities at the optical radius $V_{opt}$ spanning from $\sim 24~km\ s^{-1}$ to $\sim 300~km\ s^{-1}$. They are arranged into five optical velocity bins whose central values are given in \citet{Dipaolo.Sal.18} and whose normalized radial size is $ \sim 0.1$. 

The fourth sample, from \citet{Karukes.Sal.17} (hereafter \textbf{KS17}), consists of 36 RCs of dwarf irregular galaxies arranged into one velocity bin whose optical velocities $V_{opt}$ span from $17~km\ s^{-1}$ to $61~km\ s^{-1}$ . Its central value is $<V_{opt}>=40~km\ s^{-1}$ , and the normalized radial average size of each bin is $\sim 0.1$.

Samples 1 and 2 therefore include similar objects (normal spirals), and the second sample acts as an independent verification of the first. Samples 3 and 4 include peculiar disk systems (low surface brightness and dwarf disks). The {\bf LSB} sample is an environment extremely useful for investigating dark matter. The disk components of LSBs have a much lower surface density than normal spirals. The dwarf disks are objects with the smallest disks.

The \textbf{PSS96} and  \textbf{Ca06} samples include about 90\% of the objects belonging to  Sb-Im Hubble types. Both samples have galaxies with $V_{opt}$ in the range of normal spirals and are much larger than those in the \textbf{KS17} sample. The \textbf{LSB} includes only low surface brightness objects, which in turn are almost entirely missing in the previous three samples.
\
\subsection{Coadded rotation curves}\label{coadd}
 
  It is well known that each disk galaxy has two tags (properties) that specify it completely: \textbf{i)} $R_D$ the size of the stellar disk that  is derived  by its surface luminosity profile.
 In these objects  it is reasonable to assume that \citep{Freeman.70}
 \begin{align}
I(R)= I_0 e^{-\frac{R}{R_D.}}
 \end{align}
 
 $I_0$ is the central surface luminosity, $\sim 100\ L_{\odot}/pc^2$,
 and  $R_D=1/1.67\ R_{1/2}$ is the scale length that determines the distribution of stars in the galaxy disk, while $R_{1/2}$ is the half-light radius. The photometric quantity $R_{opt}$, analogous to the well-known half-light radius  $R_{1/2}$  in ellipticals \citep[e.g.,][]{Salucci.19}, is related to the scale length  of the Freeman disk  by 
 \begin{align}\label{roptrd}
 R_{opt}= 3.2\ R_D.   
 \end{align}
 Thus,  for exponential thin disks: $R_D =3.2/1.67\ R_{1/2}$; moreover,  $R_{opt}$ and $R_D$ are physically identical. We stress that in the case of  stellar disks with (usually mildly) nonexponential profiles,  $R_{opt}$ also still marks the dynamical edge of the disk as the quantity $3.2 \ R_D$ in the case of Freeman disks.  
 
 \textbf{ii)} Disk galaxies include irregular spirals, dwarf irregular disks and low surface brightness disks in addition to normal spirals. These types of objects are all rotationally supported, but  have very different photometric properties  \citep[see][]{Salucci.19} and therefore are  different probes for the distribution of dark matter. We specify that for a given $R_D$, the disk mass can vary largely, depending on the morphological type.

We can represent all the rotation curves of disk systems by means of a universal profile \citep[e.g.,][]{Salucci.Lap.07}. In detail, the universal RC (URC) is obtained from thousands of individual RCs when they are arranged in a number of coadded\textup{} rotation curves. The uncertainties of coadded (stacked) rotation curves are significantly smaller than those of the individual RCs. Moreover, these coadded curves are as extended  as the most extended individual RCs: $r_{outermost}>5 R_D$, where $r_{outermost}$ is the center of the outermost radial bin in each coadded RC and has a very high spatial resolution ($< 1/4 \ R_D$). 
The process of building such curves is explained in detail in the original works, but we briefly recall it here. 

The basic assumption underlying the coaddition process and verified in disk systems is that the galaxies with similar $V_{opt}\equiv V(R_{opt})$  (that also have  similar $R_{opt}$) have similar rotation curves, especially when the latter are expressed in the normalized radial coordinate $x\equiv R/R_{opt}$  \citep{ Salucci.Yeg.07, Gammaldi.Kar.18, Fune.18, Rhee.97}. This leads to similar mass distributions, with the DM density always taken as a cored distribution.

The concept of coadded RCs, implicit in \citet{Rubin.Bur.85}, which was pioneered
by \citet{Persic.Sal.91}, set by \citet{PSS96}, and extended to large galactocentric radii and different Hubble types
by \citet{Salucci.Lap.07}, \citet{Dipaolo.Sal.18}, and \citet{Karukes.Sal.17} allows us to derive $V_{coadd}(R/R_{opt},M_I)$ or $V_{coadd}(R/R_{opt},V_{opt})$, that is, the average (coadded) RCs of galaxies with the magnitude lying in a bin centered at the I-band absolute magnitude $M_I$ and of size $\delta M_I$ (or  centered at $ V_{opt}$ and of  size $\delta V_{opt}$). The number of objects and the position and size of each magnitude (or velocity) bin are indicated in detail in the original works. 
\
The coadded RCs were built in three steps in a process that is  similar to the process used in cosmological numerical simulations to work out the averaged DM  density distribution of   halos of  similar virial mass $M_{vir}$ \citep[see:][]{NFW96}. We describe here  the details for one of the four  samples  we used (\textbf{PSS96}); the procedure is only marginally different from those adopted for the other three samples.

\textbf{i)}
 The whole I-magnitude range is divided into 11 successive
bins, each of which is centered at $M_{\textbf{j}, I}$ ($\textbf{j}=1....,11$) and  has a size $\delta M_{\textbf{j},I}$. The 11 values of the latter quantities, alongside with the 11 values of  $  V_{\textbf{j}, opt}$, $R_{\textbf{j}, opt}$ , are given in Table \ref{rePSS} \citep[also in][]{PSS96}.  The rotation curve $V_{\textbf{jk}}(R)$ of each galaxy of the sample assigned to its corresponding $\textbf{j}$ -magnitude  bin is normalized by its $V_{\textbf{jk}, opt}$ value. Next, by using the known galaxy  value of  $R_{\textbf{jk}, opt}$,  each  RC is expressed in terms of its normalized radial coordinate 
$x_{\bf jk}\equiv R/R_{\textbf{jk}, opt}$. Here, $_{\bf j}$ is the index of the magnitude (velocity) bins, $_{\bf jk}$ is the index of the various galaxies in the above bins before coaddition, $_{\bf ji}$ is the index of the radii at which  $_{\bf j}$ {\it coadded} RC have measurements. We directly determined in the first sample (\textbf{PSS96}) the various $R_{{\bf jk}, opt}$  from the definition, for \textbf{LSB} and \textbf{KS17}, the Freeman disk is an excellent distribution for their stellar content, and this also holds for the spirals in \textbf{Ca06}, if slightly less so. In the above works, the  $R_{{\bf jk}, D}$ values were used to derive $R_{{\bf jk}, opt}$ from Eq. (\ref{roptrd}).  In this way, every RC of our sample was then double-normalized and tagged as  
\begin{equation}
v_{{\bf jk}}(x_{{\bf jk}})\equiv \frac{V_{\textbf{jk}}(x_{\bf jk})}{V_{\textbf{jk}, opt}}
.\end{equation}
Let us notice that $V$ is  velocity in physical units while $v$ is double-normalized velocity.  
\textbf{ii)}   Each  of the $N_{\bf j}$   double-normalized{ \it \textup{individual}}   rotation curves  $v_{\textbf{jk}}$ in each $\textbf{j}$  bin was binned  into 20 radial bins  of length 0.1 and index ${\bf ji,}$ and the number of data $N_{{\bf ji}}$ was then averaged to obtain the double-normalized  coadded RCs and their uncertainties computed from their r.m.s  $\delta v_{{\bf j}, coadd}(x_{\textbf{ji}}, M_{ \textbf{j}, I}) $ and $N_{{\bf ji}}$:
$\delta v_{{\bf j}, coadd}(x_{\textbf{ji}}, M_{ \textbf{j}, I})/  (N_{\bf ji}-2)^{1/2}$.

 \textbf{iii)} The final step is that the double-normalized curves are denormalized using the averaged values of $V_{{\bf jk},opt}$ and $R_{{\bf jk},opt}$.

In general,  the extension of the 26 coadded RCs is $x_{{\bf j}, outermost}= 1.75\pm 0.25  $  and the positions of the various $x_{\bf ji}$ are given in the online table for 
${\bf{j}}=1...26$. For $x>1.2$, the concurring  RCs are mostly  HI RCs. We recall that \citet{Gentile.Sal.04} have shown that HI and optical RCs describe the same gravitational potential. Noticeably, these  coadded curves vary very little: almost always, $\delta V_{coadd} \leq 0.03 \ V_{coadd}$ \citep[e.g.,][]{Salucci.19} because for most of  the radial bins in each coadded RC, we have many measurements, from $20$ to $200$. When these are averaged,  the statistical component of the variance  in the velocity measurements is much reduced. Furthermore,  $v_{\textbf{jk}}(x_{\bf jk},M_{{\bf j}, I}),$   the individual double-normalized RCs of objects with the same magnitude $M_I$ are all very similar.
 
  It is also important to stress in reference to all samples (with $\textbf{j}$= 1....26),  that in  each of the $\textbf{j}$ luminosity (velocity) bins, the values of the normalization velocities $V_{\textbf{jk}, opt}$ of the individual RCs assigned to the $\textbf{j}$ bin  are very similar. They each differ from the averaged bin value $V_{\textbf{j}, opt}$ by less than 10\%. For simplicity of notation for the quantities  $V_{\textbf{j}, opt}$ and $R_{\textbf{j}, opt}$ we do not indicate the  sign <>, but these quantities must be averaged over the index $\textbf{jk}$ with  fixed $\textbf{j}$. In the first luminosity (velocity) bin of each sample, the values of the self-normalization velocities $V_{\textbf{jk}, opt}$ may vary significantly among the galaxies of the bin; however, in these cases, corresponding to the lowest luminosity (velocity) bins of the four samples, the profiles of the individual RCs, that is, $d\log V/d\log R$, which are crucial for the mass modeling, are all very similar and independent of  $V_{\textbf{jk}, opt}$, for example. We stress that the averaged value  $<d\log V/d\ log R>$ between $1\ R_D$ and $3\ R_D$ is lower than $0.02$ for
all the  26 coadded RCs,  which is   irrelevant in the mass modeling  process. 

Similarly, in  each of the 26 magnitude (velocity) $\textbf{j}$ bins  the   radial normalization quantities $R_{\textbf{jk}, opt}$ (fixed ${\bf j}$) are very similar, and the r.m.s. of  the averaged values $R_{\textbf{j}, opt}$ is   $<25\%$. Moreover, because the individual RCs locally have  a mostly linear profile, $V(R +\delta R) \simeq V(R) + const. \times \delta R$, the variance of the various $R_{\textbf{jk}, opt}$(fixed ${\bf j}$) with respect to the average $R_{\bf j, opt}$  affects the coaddition process only very mildly. Therefore, the finiteness of the 26 luminosity (velocity) bins does not affect the structure of the  the double-normalized coadded RCs and their subsequent denormalizations.

The URC of disk systems  is instead the {\it \textup{analytical}} function devised to fit the coadded RCs $V_{{\bf j}, coadd} (r,M_{\textbf{j}, I})$ in \citet{PSS96} and  in the others in the subsequent works. It was chosen as the sum in quadrature of the components to the circular velocity coming from a Freeman stellar disk and a Burkert dark matter halo \citep{Bur.95,PSS96}. In this work,  the URC  provides a crucial model comparison. In previous works, we have shown that all the 26 coadded RCs that we used here, are perfectly fit by this function, that is, in more detail, by a velocity model including a {\it \textup{cored}} DM halo with $\rho_{DM} \propto ((R+r_0)(R^2+r_0^2))^{-1}$  and a Freeman disk. This model has the same number of free parameters as the model we analyze here: the core radius $r_0$, the central DM density $\rho_0$  , and the disk mass $M_D$. In all 26 cases the URC fitting uncertainties were found to be lower than the r.m.s of the $V_{{\bf j}, coadd}(x_{\textbf{ji}}, M_{\textbf{j}, I}\ or \  V_{\textbf{j}, opt})$ measurements \citep{PSS96, Catinella.Gio.06, Dipaolo.Sal.18, Karukes.Sal.17, Salucci.Lap.07}.

 We briefly recall that at a physical radius $R$, the normalized radius $x$ and the double-normalized circular velocity $v(x)$ are defined as
 
 \begin{equation}
 x\equiv \frac{R}{R_{opt}} \ \ \ \ \ \ \ \ \ \ \ \ \ \ \  v(x)\equiv \frac{V(x)}{V_{opt}}
.\end{equation}
 
For a double-normalized {\it \textup{coadded}} RC {\bf j} with velocity data at $x_{\bf ^{ji}}$, with ${\bf j}=1,....26$, we have 

 \begin{equation}
 x_{\bf {ji}}\equiv \frac{R_{\bf {ji}}}{R_{\bf{j}, opt}} \ \ \ \ \ \ \ \ \ \ \ v(x_{\bf {ji}})\equiv \frac{V(x_{\bf {ji}})}{V_{\bf{j}, opt}}
,\end{equation}
which gives $R_{\bf ji} $,  the positions  in physical units  of the coadded velocity data of $v_{\bf j}(x)$.
 
\begin{figure*}[ht!]
\begin{center}
\includegraphics[scale=1]{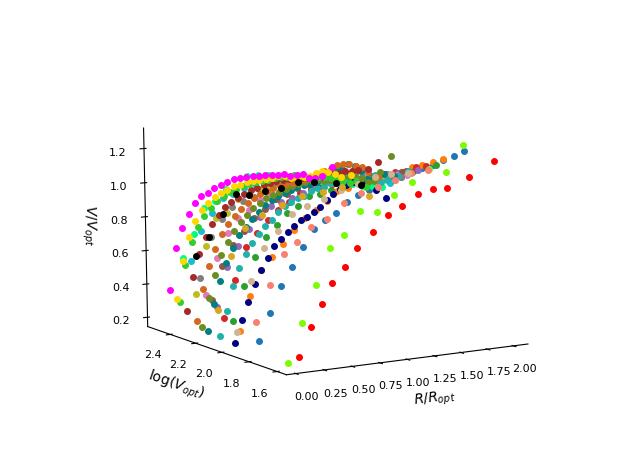}
\end{center}
\caption{$R/R_{opt}-\log{V_{opt}}-V/V_{opt}$ for the 26 coadded RCs. The legend is the same as in Fig. \ref{obdat}. For clarity we do not plot the error related to each measurement; this can be found in the online table.}
\label{obdat2}
\end{figure*}

 Fig. \ref{obdat} shows the 26 coadded RCs in double-normalized coordinates. To express them in physical units, it is necessary to rescale them according to the above relations and the values in Tables (\ref{rePSS}-\ref{reCa}). We also plot the coadded RCs in a 3D form to show that in the $(v,x,V_{opt})$ coordinates system, they clearly define a 3D surface that leads to the concept of the URC (see Fig. \ref{obdat2}).

\subsection{Mass model}\label{mmod}

In the $\Lambda$CDM scenario, we have from  the N-body simulations that the virialized dark halos show a universal spherically averaged density profile, 
\begin{align}\label{nfwden}
\rho(R)=\frac{\rho_s}{(R/r_s)(1+R/r_s)^2},
\end{align}
where $R$ is the radial spherical coordinate, and  $\rho_s$ and $r_s$ are the characteristic density and the scale radius of the dark halo. It is useful to define the concentration $c\equiv R_{vir}/r_s$ , where $R_{vir}$ is the virial radius given by $M_{vir}\equiv 4/3\pi\ 100\rho_{crit}R^3_{vir}$. The critical density is $\rho_{crit}=3H_0^2/(8\pi G)$ with $H_0$ the current value of Hubble's parameter.  We assume $H_0= 72~ km\ s^{-1}Mpc^{-1}$. No  result of this paper changes by assuming the lower value of $H_0=67~km\ s^{-1}Mpc^{-1}$ favored by Plank CMB measurements \citep{Planck18}.
\
In the circular velocity model, the contribution of the DM halo to the circular velocity adds in quadrature to that of the stellar disk as
 
\begin{align}\label{urcv}
V^2_{mod}(R)=V^2_{D}(R)+V^2_{NFW}(R),
\end{align}
where $V_D$ and $V_{NFW}$ are the contribution of the stellar Freeman disk and the dark halo, respectively. In the disk systems we consider in this work, the stellar component is described by an exponential Freeman  disk \citep[see][]{Freeman.70} with a surface density given by
$$
\Sigma_D(R)=\frac{M_D}{2\pi R_D^2}e^{-\frac{R}{R_D}}=\left[\frac{M_D}{2\pi (R_{opt}/3.2)^2}\right]e^{- 3.2 x. }
$$

Its contribution to the circular velocity is
\begin{align}\label{vd0}
V_D^2(x)=\frac{1}{2}\frac{G \ M_D}{R_D}(3.2x)^2\left[I_0(1.6x)K_0(1.6x)-I_1(1.6x)K_1(1.6x)\right],
\end{align}
where it is worth  recalling that $M_D$ is the disk mass, $x\equiv R/R_{opt}$ and $I_n, K_n$ are the modified Bessel functions of $n^{th}$ order.

The dark halo contribution to the circular velocity $V_{NFW}^2=G \ M_{NFW}(R) /R$ that we investigate here is the contribution generated by the NFW dark halo density profile ($\rho_{NFW}$) and is given by 
\begin{align}\label{VNFW}
V_{NFW}^2(R)=\frac{GM_{vir}}{R}\frac{\ln(1+c\ y)-\frac{c\ y}{1+c\ y}}{\ln(1+c)-\frac{c}{1+c}},
\end{align}
where $M_{vir}$ is the halo virial mass, $y=R/R_{vir}$ is the  normalized radial coordinate, and $c=R_{vir}/r_s$ is the concentration parameter.

The velocity model that we tested with 26 coadded rotation curves obtained from more than 3200 individual RCs that consist of  more than $10^5$ independent kinematical measurements is  

\begin{equation}\label{modv}
V^2_{mod}(R;M_D,c, M_{vir})=V^2_D(R; M_D)+V^2_{NFW}(R;c,M_{vir})
.\end{equation}

the notation $(c,M_D,M_{vir})$ has to be considered as a  $26 \times 3 $ matrix. 

Finally, we stress that we did not test an empirical dark matter density profile  \citep[as in][]{Bur.95} for which we would need data out to the galaxy virial radius to prove its success,  but a theoretical profile obtained from  N-body simulations in the $\Lambda$CDM scenario, whose functional form is believed to be perfectly known from $r=0$ to $r=R_{vir}$. In detail, the two free parameters $\rho_s$ and $r_s$ can be derived by the the inner regions of the RCs, that is, for $R<r_s$ (see Eq. \ref{nfwden}). We do not need very extended kinematics to disprove the presence of an NFW halo.
It is clear that with more external data the estimate of these parameters may change, but our aim is not to show a global failure of the NFW + disk model but only its extremely serious troubles in the region in which the model-independent analysis of RCs has revealed DM density cores,  that is, for $r<r_0 \sim r_s$. We did not investigate whether for $r> R_{opt}$ the DM halo density converges to an NFW, as it appears to do in spirals \citep{Salucci.Lap.07}.
\
\section{Results}\label{result}
The data fitting was performed by means of the nonlinear least-square method ($\chi^2$-Levenberg-Marquardt method). We fit each coadded rotation curve with the above model. The $\chi^2$ statistic is defined as

\begin{equation}
\chi^2_{\bf j}= \sum_{{\bf i}=1}^{N_{\bf ji}}\left(\frac{\left(V_{{\bf ji},coadd}-V_{{\bf ji},mod}(X)\right)^2}{\sigma^2_{{\bf ji},coadd}N_{{\bf ji}}}\right)
,\end{equation}

where $V_{{\bf ji}, coadd}$ is the value of coadded RCs at  $x_{\bf ji}$ tagged as  ${\bf j}$ ; same definition stands for $\sigma_{{\bf ji}, coadd}$ and $V_{{\bf ji}, mod}$ that are the   (observational) uncertainties of the ${\bf j}$ coadded RCs value  at $x_{\bf ji}R_{{\bf j}, opt}$ and the corresponding model  value, respectively. $N_{\bf ji}$ are the number of kinematical  measurements of each coadded RC (see Tables \ref{rePSS}-\ref{reCa}), and $X$ is the set of three free parameters of the model, $X=(M_D,M_{vir},c), $ which are to be determined through the fitting procedure. The values of $ R_{\bf ji}=x_{\bf ji} R_{{\bf j}, opt}$, $V_{{\bf ji}, coadd}$,  and $\sigma_{ \bf ji}$ are given in an online file. 

These best-fit values and those  of the reduced $\chi^2_{\bf j}$ ($\chi^2_{{\bf j}, red}$), where $\chi^2_{{\bf j }, red}=\chi^2_{\bf j}/(N_{{\bf ji}}-3)$ and the denominator is the number of degrees of freedom of the model,  are shown in Tables (\ref{rePSS}-\ref{reCa}) for the 25 coadded RCs investigated here. In Fig. \ref{trg22} we show the triangular plots for 25 mass models. 
 
 The coadded RC of \textbf{KS17} has been modeled in the original paper, where the authors found
\begin{align}
\log M_{vir}=11.68\pm 0.87, \ \ 
 c=4.73\pm 3.19, \ \ 
 \log M_D=2.5_{-2.5}^{+2.5},
\end{align}
with $\chi^2_{red}\approx 12$.

 At this point, we state that in order to reclaim success, we expect from the model under investigation that it fits the data satisfactorily ($\chi^2_{red}\simeq 1$) and that the values of free parameters are at least in a fair agreement with the following relations holding for 
 disk galaxies: $M_D= G^{-1} \alpha V_{opt}^2 R_{opt}$ with $0.2<\alpha<0.8 $ from the colors and the luminosities of spirals \citep[e.g.,][] {Salucci.Yeg.08}, $c\sim (7-17)$ from N-body simulations \citep{Bullock.Kol.01} according to the halo masses (see Eq. \ref{cm}), and $M_{vir}\sim 10^{12} M_\odot (V_{opt}/(200~ km\ s^{-1}))^3$ from weak-lensing measurements  and  abundance-matching technique \citep[e.g.,][]{Salucci.19}.
 
 Overall, the model performs very poorly. The most relevant problems are that there is  a wide range of values for the concentration parameter $c$ , which ranges  from $1.5$ to $28.3,$ and as indicated in boldface in Tables (\ref{rePSS}-\ref{reCa}), in $50 \%$ of the cases it is well outside the simulation range. 
 
 There are also 12  entirely implausibly high values for the halo virial masses in  the 26 best-fitting models. We stress that these high values do not 
originate from the lack of kinematical measurements in the outer parts of galaxy halos (i.e., for $R/R_{opt}>1$ ), but from the values of the (coadded) RCs in the innermost regions of the galaxies ($R< R_{opt} $ ): the RC profiles are so steep (i.e., $V_{coadd} \propto R$) there   that in order to reproduce this RC characteristic with an NFW halo profile, the best-fit model is forced to have very high values of the  parameters $r_s$ and $\rho_s$, which in turn yields implausibly high values for the  virial halo mass, for which  $M_{vir} \propto \rho_s\ r_s^3$.

 Moreover, according to Tables (\ref{rePSS}-\ref{reCa}), in order to minimize the $\chi_{red}^2$ value, the best fit (NFW halo + Freeman disk) model forces the parameter $M_D$ in 5 cases out of
the total 26 to assume  values that are far lower than those obtained from the $L_B$ galaxy luminosities  and average spiral mass-to-light ratios  $M_D/L_B=2$ . The uncertainties on $M_D$ propagated  from those in $R_{opt}$ are smaller than $25 \%$ and therefore negligible for the results of this work. From its  maximum value, obtained with the assumption that at $R= 2.2\ R_D$  the disk component entirely dominates  the circular velocity $V(2.2\ R_D)$,  we have $M_{D, max} \sim G^{-1} V^2\ (2.2\ R_D) \ 2.2\ R_D$. The actual disk mass  can be a fraction of it.

 Furthermore, the $\chi^2_{{\bf j}, red}$ values are  not satisfactory. In detail,  we find that this quantity exceeds the value of  1.4 in  16 cases.  We recall that the URC velocity model (Burkert halo + Freeman disk) fits all the 26 coadded RCs we used here extremely  well \citep[see][]{PSS96, Lapi.Sal.18, Dipaolo.Sal.18, Karukes.Sal.17}. However, poor fits are only a part of the failure of the model we tested. 
 
\begin{table*}
\begin{center}
\caption{Values of the best-fit parameters of the mass modeling with their $1\sigma$ uncertainties for the \textbf{PSS96} sample. Column $(1)$ lists the number of the bins,  Col. $(2)   $ the I-band luminosity magnitude, Col. $(3)$ the number of data points in each bin, Col. $(4)$ the optical radius, Col. $(5)$ the optical velocity, Col. $(6)$ the concentration, Col. $(7)$ the disk mass, Col. $(8)$ the halo mass (virial mass), and Col. $(9)$ the reduced $\chi^2$ value. The $\chi_{red}^2$ and the values of the free parameters in boldface mean that they failed the test, and values in red indicate that this quantity is not sufficiently well estimated.}\label{rePSS}
\begin{tabular}{ccccccccc}
\hline
 $j$ & $M_{j, I}$ & $N_{ji}$ & $R_{j, opt}$ & $V_{j, opt}$ & $c$ & $\log (M_D)$ & $\log(M_{vir})$ & $\chi^2_{j, red}$ \\[0.8ex]
 $- $ & $ -$ & $ -$ & $(kpc)$ & $(km\ s^{-1})$ & $- $ & $(M_\odot )$ & $(M_\odot) $ & $- $ \\[0.8ex]
$\textbf{(1)}$ & $\textbf{(2)}$ & $\textbf{(3)}$ & $\textbf{(4)}$ & $\textbf{(5)}$ & $\textbf{(6)}$ & $\textbf{(7)}$ & ${\textbf{(8)}}$ & ${\textbf{(9)}}$\\ \hline
$1$ & $-18.5$ & $20$ & $4.6$ & $75$ & $\textcolor{red}{\textbf{2.5}^{+\infty}_{-\infty}}$ & $\textcolor{red}{8.23^{+0.35}_{-\infty}}$ & $\textbf{13.2}^{+1}_{-0.8}$ & $\textbf{4.8}$\\ \hline
$2$ & $-19.4$ & $20$ & $5.7$ & $104$ & $11.1^{+2}_{-3}$ & $\textcolor{red}{8.77^{+0.5}_{-\infty}}$ & $\textbf{11.9}^{+0.3}_{-0.14}$ & $\textbf{3.7}$ \\ \hline
$3$ & $-20.0$ & $20$ & $6.5$ & $116$ & $8.8^{+1.8}_{-3.2}$ & $9.7^{+0.1}_{-0.2}$ & $\textbf{12.0}^{+1.8}_{-0.24}$ & $\textbf{5.1}$ \\ \hline
$4$ & $-20.5$ & $20$ & $7.6$ & $135$ & $8.5^{+3.2}_{-3.2}$ & $9.98^{+0.1}_{-0.12}$ & $\textbf{12.2}^{+0.5}_{-0.25}$ & $\textbf{3.7}$ \\ \hline
$5$ & $-20.9$ & $20$ & $8.9$ & $154$ & $\textcolor{red}{10.9^{+\infty}_{-4.8}}$ & $9.6^{+0.36}_{-0.1}$ & $\textbf{12.3}^{+0.58}_{-0.16}$ & $\textbf{2.8}$ \\ \hline
$6$ & $-21.2$ & $20$ & $10.1$ & $169$ & $10.3^{+1.9}_{-2}$ & $10.21^{+0.08}_{-0.1}$ & $12.3^{+0.18}_{-0.11}$ & $1.1$ \\ \hline
$7$ & $-21.6$ & $20$ & $11.5$ & $185$ & $9.9^{+2.7}_{-2.9}$ & $10.53^{+0.06}_{-0.07}$ & $12.2^{+0.26}_{-0.14}$ & $0.5$ \\ \hline
$8$ & $-22.0$ & $20$ & $13.5$ & $205$ & $\textbf{25.7}^{+1.7}_{-1.8}$ & $10.3^{+0.13}_{-0.19}$ & $11.9^{+0.04}_{-0.04}$ & $0.9$ \\ \hline
$9$ & $-22.2$ & $20$ & $15.3$ & $225$ & $\textbf{15.6}^{+4.3}_{-8.5}$ & $10.78^{+0.18}_{-0.15}$ & $12.1^{+0.35}_{-0.06}$ & $1.3$ \\ \hline
$10$ & $-22.6$ & $20$ & $18.0$ & $243$ & $\textbf{28.3}^{+0.9}_{-0.9}$ & $10.62^{+0.06}_{-0.07}$ & $12.1^{+0.01}_{-0.02}$ & $0.2$ \\ \hline
$11$ & $-23.2$ & $20$ & $22.7$ & $279$ & $\textbf{19.3}^{+3.6}_{-4.3}$ & $11.2^{+0.09}_{-0.09}$ & $12.2^{+0.04}_{-0.04}$ & $0.5$ \\ \hline
\end{tabular}
\end{center}
\end{table*}

\begin{table*}
\begin{center}
\caption{Same as in Table \ref{rePSS} for the \textbf{LSB} sample.}\label{reLSB}
\begin{tabular}{ccccccccc}
\hline
 $j$ & $M_{j, I}$ & $N_{ji}$ & $R_{j, opt}$ & $V_{j, opt}$ & $c$ & $\log (M_D)$ & $\log(M_{vir})$ & $\chi^2_{j, red}$ \\[0.8ex]
 $- $ & $ -$ & $ -$ & $(kpc)$ & $(km\ s^{-1})$ & $- $ & $(M_\odot) $ & $(M_\odot) $ & $- $ \\[0.8ex]
$\textbf{(1)}$ & $\textbf{(2)}$ & $\textbf{(3)}$ & $\textbf{(4)}$ & $\textbf{(5)}$ & $\textbf{(6)}$ & $\textbf{(7)}$ & $\textbf{(8)}$ & ${\textbf{(9)}}$\\ \hline
$12$ & $...$ & $12$ & $5.5$ & $44$ & $\textcolor{red}{\textbf{1.5}^{+\infty}_{-\infty}}$ & $8.5^{+0.25}_{-0.6}$ & $\textcolor{red}{\textbf{12.5}^{+\infty}_{-0.4}}$ & $\textbf{4.0}$ \\ \hline
$13$ & $...$ & $12$ & $6.9$ & $73$ & $\textcolor{red}{8.5^{0.2}_{-\infty}}$ & $\textcolor{red}{9.01^{0.08}_{-\infty}}$ & $11.4^{0.01}_{0.06}$ & $\textbf{5.9}$ \\ \hline
$14$ & $...$ & $12$ & $11.8$ & $101$ & $\textbf{3.4}^{+1.8}_{-2.4}$ & $9.93^{+0.08}_{-0.1}$ & $\textbf{12.3}^{+1.1}_{-0.36}$ & $1.1$ \\ \hline
$15$ & $...$ & $9$ & $14.5$ & $141$ & $\textbf{12.7}^{+2.1}_{-2.4}$ & $10.31^{+0.1}_{-0.1}$ & $11.5^{+0.08}_{-0.08}$ & $\textbf{1.5}$ \\ \hline
$16$ & $...$ & $11$ & $25.3$ & $206$ & $\textcolor{red}{\textbf{23.4}^{+\infty}_{-1.9}}$ & $\textcolor{red}{\textbf{7.0}^{+3.4}_{-\infty}}$ & $12.1^{+0.04}_{-0.06}$ & $\textbf{2.26}$ \\ \hline
\end{tabular}
\end{center}
\end{table*}

\begin{table*}
\begin{center}
\caption{Same as in Table \ref{rePSS} for the \textbf{Ca06} sample.}\label{reCa}
\begin{tabular}{ccccccccc}
\hline
 $j$ & $M_{j, I}$ & $N_{ji}$ & $R_{j, opt}$ & $V_{j, opt}$ & $c$ & $\log (M_D)$ & $\log(M_{vir})$ & $\chi^2_{j, red}$ \\[0.8ex]
 $- $ & $ -$ & $ -$ & $(kpc)$ & $(km\ s^{-1})$ & $- $ & $(M_\odot )$ & $(M_\odot) $ & $- $ \\[0.8ex]
$\textbf{(1)}$ & $\textbf{(2)}$ & $\textbf{(3)}$ & $\textbf{(4)}$ & $\textbf{(5)}$ & $\textbf{(6)}$ & $\textbf{(7)}$ & $\textbf{(8)}$ & ${\textbf{(9)}}$\\ \hline 
$17$ & $-19.4$ & $23$ & $5.7$ & $104$ & $11.6^{+2.7}_{-4.8} $ & $\textcolor{red}{9.15^{+0.29}_{-\infty}}$ & $11.7^{+0.5}_{-0.2}$ & $\textbf{3.0}$ \\ \hline
$18$ & $-20.45$ & $29$ & $7.6$ & $135$ & $12.9^{+3.3}_{-3.2} $ & $9.68^{+0.18}_{-0.35}$ & $\textbf{12.0}^{+0.22}_{-0.14}$ & $\textbf{1.8}$ \\ \hline
$19$ & $-21.25$ & $29$ & $10.1$ & $169$ & $10.1^{+3}_{-3} $ & $10.17^{+0.11}_{-0.17}$ & $12.4^{+0.35}_{-0.19}$ & $\textbf{2.2}$ \\ \hline
$20$ & $-21.57$ & $30$ & $11.5$ & $185$ & $8.5^{+3.2}_{-3} $ & $10.48^{+0.12}_{-0.1}$ & $\textbf{12.5}^{+0.42}_{-0.22}$ & $\textbf{2.1}$ \\ \hline
$21$ & $-21.96$ & $29$ & $13.5$ & $205$ & $7.3^{+3}_{-2.8} $ & $10.67^{+0.09}_{-0.09}$ & $\textbf{12.8}^{+0.51}_{-0.28}$ & $\textbf{1.6}$ \\ \hline
$22$ & $-22.64$ & $30$ & $18.0$ & $243$ & $\textcolor{red}{\textbf{2.0}^{0.9}_{-\infty}}$ & $11.11^{+0.01}_{-0.01}$ & $\textcolor{red}{\textbf{14.0}_{0.4}^{+\infty}}$ & $0.3$ \\ \hline
$23$ & $-23.19$ & $30$ & $22.6$ & $279$ & $\textbf{18.3}^{+1.2}_{-1.2} $ & $11.05^{+0.05}_{-0.05}$ & $\textbf{12.4}^{+0.01}_{-0.01}$ & $1.0$ \\ \hline
$24$ & $-23.40$ & $28$ & $24.7$ & $293$ & $\textbf{22.3}^{+1.5}_{-1.5} $ & $10.97^{+0.1}_{-0.1}$ & $12.4^{+0.02}_{-0.02}$ & $\textbf{2.0}$ \\ \hline
$25$ & $-23.80$ & $26$ & $29.1$ & $330$ & $\textbf{26.8}^{+0.8}_{-0.9} $ & $10.99^{+0.1}_{-0.1}$ & $12.5^{+0.02}_{-0.02}$ & $1.0$ \\ \hline
\end{tabular}
\end{center}
\end{table*}

The performance of the model under investigation was also tested by correlating the best-fit values of the two DM structural parameters $c$ and $M_{vir}$ and  then comparing them  with the corresponding scaling relation that emerges in  the N-body simulations within the $\Lambda$CDM scenario \citep[][]{Bullock.Kol.01}:
\begin{align}\label{cm}
c=13.6\left(\frac{M_{vir}}{10^{11}M_\odot}\right)^{-0.13},
\end{align}

\begin{figure}[t!]
\begin{center}
\includegraphics[scale=0.6]{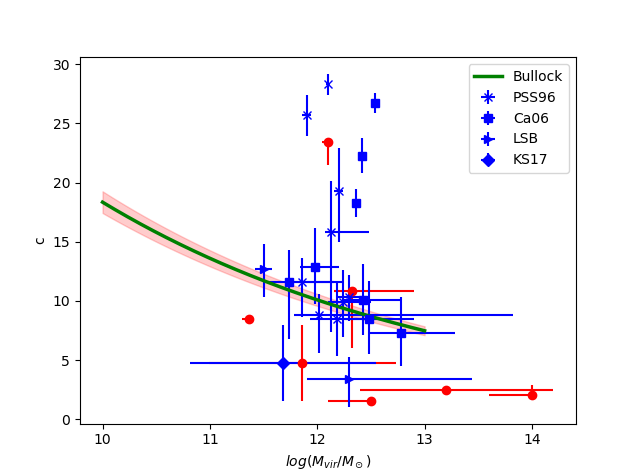}  
\end{center}
\caption{Relation between $c$ and $\log M_{vir}$ from the 26 mass models. Blue points indicate the results of this paper (red points indicate a very large uncertainty on their estimates). The $\Lambda$CDM outcome from simulations is also shown (green line).}
\label{cvrelat}
\end{figure}

 The cosmic variance in $\log c$ is 0.1 dex, which is negligible for the aim of  our work. Fig. \ref{cvrelat} shows the $c-\log(M_{vir})$ relation of our 26 best-fit models  with their corresponding uncertainties. We can easily realize that most of the best-fit values  lie very far from the  Eq. (\ref{cm}) relation, of which they are unable to trace even the gross trend.

\begin{figure}[t!]
\begin{center}
 
\includegraphics[scale=0.6]{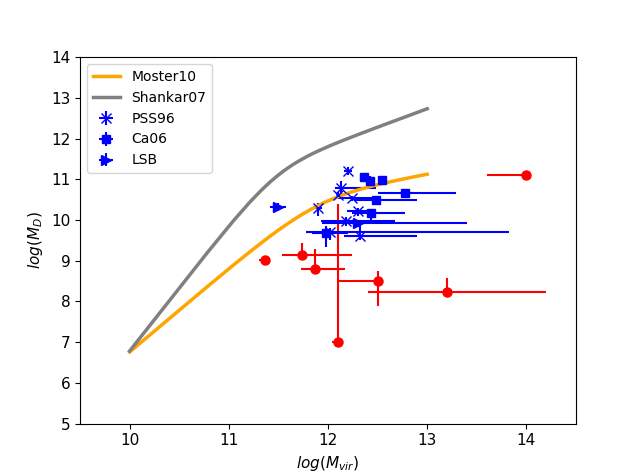}  
\end{center}
\caption{ $M_D- M_{vir}$ relationship in logarithmic scale obtained from  the 25 coadded RCs (blue points). The red circles are the same as in Fig. \ref{cvrelat}. The orange and gray lines correspond to relations (\ref{sd1}) and (\ref{sa}). The value of $\log M_D$ for \textbf{KS17} has the  extremely  discrepant coordinates (see the text).}
\label{dvrelat}
\end{figure}

Another scaling relationship of great importance is the disk mass $M_D$ \textit{} versus the virial halo mass $M_{vir}$; noticeably,  this relation  can be independently derived  by the abundance-matching method \citep[e.g.,][]{Moster.Som.10}. We find

\begin{align}\label{sd1}
\frac{M_D}{M_{vir}}= A \left[\left(\frac{M_{vir}}{M_1}\right)^{-\beta}+\left(\frac{M_{vir}}{M_1}\right)^\gamma\right]^{-1},
\end{align}
where $M_1, A, \beta, \text{and } \gamma$ are 
\begin{align}
\log M_1=11.884;\ A=0.0564;\ \beta=1.057;\ \gamma=0.556.
\end{align}

A completely independent analysis  has yielded the disk mass $M_D$ \textit{} versus virial halo mass $M_{vir}$ (see \citet{Shankar.Lap.06}). This is   just  mildly  different  from the mass in Eq. (\ref{sd1}), also considering that both have an uncertainty of 0.2 dex in $\log (M_D$), 

\begin{align}\label{sa}
M_D=2.3\times 10^{10}M_\odot\frac{\left(\frac{M_{vir}}{3\times 10^{11}M_\odot}\right)^{3.1}}{1+\left(\frac{M_{vir}}{3\times 10^{11}M_\odot}\right)^{2.2}},
\end{align}
 which we  also included in the comparison with the best-fit values of the model under investigation.
 The stellar-halo mass relationship  so obtained is shown in Fig. \ref{dvrelat}, along with the above independent relationships (\ref{sd1}) and (\ref{sa}). It is evident that most of the values of the (NFW + stellar disk) model lie very far from either relation. The results of \textbf{KS17} are   entirely inconsistent with the observational relations.

The poorest performance of  the  model under study  is probably  related to the fact that this  never shows a halo mass with $M_{vir}< 10^{12} \ M_\odot$. Halo masses like this  are known to  exist in a great number in the $\Lambda CDM$ scenario.

The failure of the model under analysis to reproduce the kinematics of spirals is summarized  in Tables (\ref{rePSS}-\ref{reCa}) and Figs. \ref{cvrelat} and \ref{dvrelat}. In order to have a more complete view of the problem, we plot the values of three important structural parameters derived from the best fits of the 26 coadded RCs:
\begin{figure*}[t!]
\begin{center}
\qquad
\vskip 0.4cm
\includegraphics[scale=0.18]{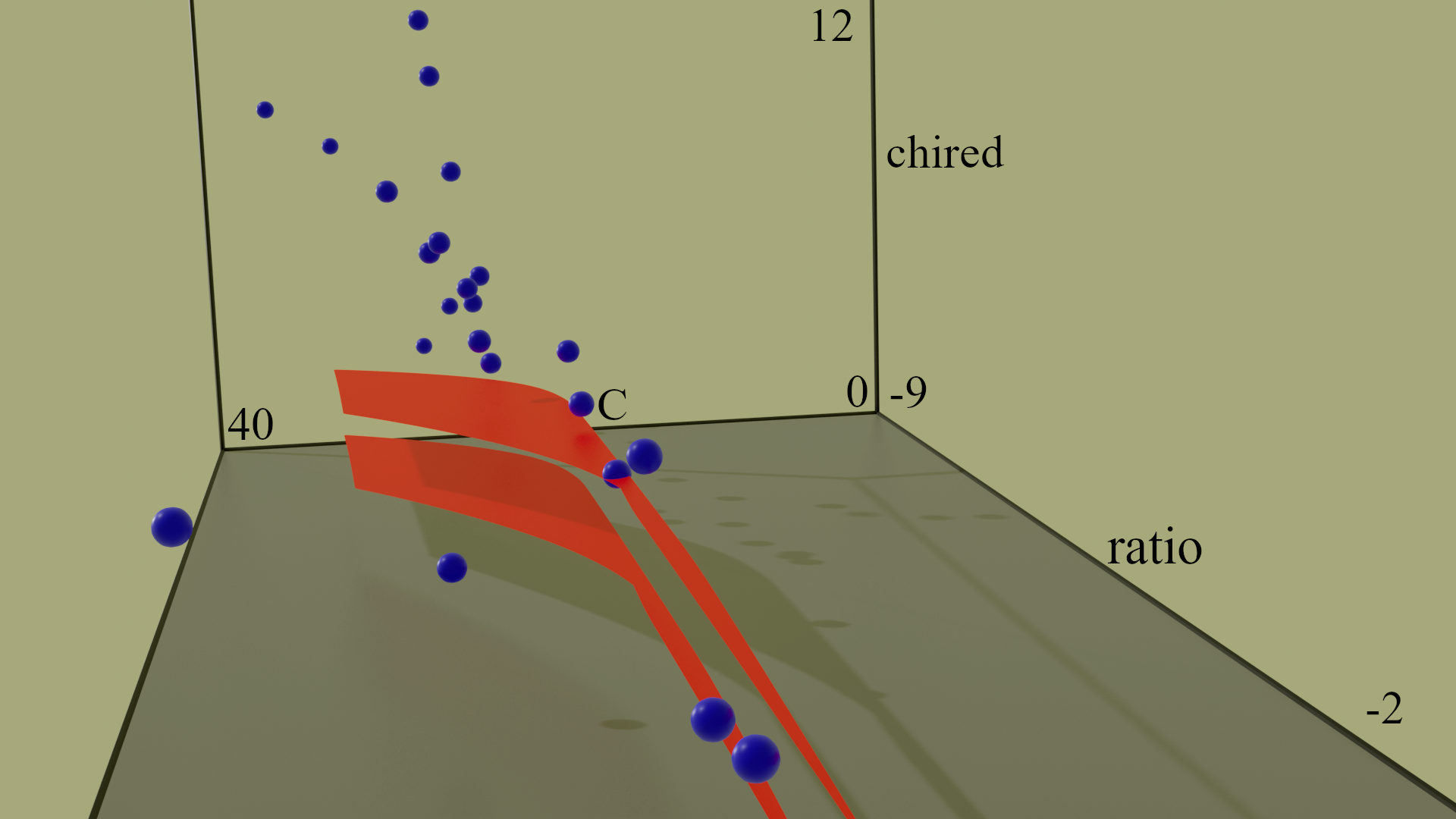}
\caption{ Relation between $c$, ratio and chired.  $ratio \equiv \log (M_D/ M_{vir})$ and $chired \equiv \chi^2_{red}$. The values are obtained from the 26 mass models (3D blue spheres) and the $\Lambda CDM$ expectation (the volume between two red surfaces).}
\label{3dplt}
\end{center}
\end{figure*}
in Fig. \ref{3dplt} we show $c$ as a function of $\log \ (M_D/M_{vir})$ and $\chi^2_{red}$. The two surfaces restrict the volume in which we expect the values of the DM parameters to lie in, which  means that the  $\Lambda$CDM NFW model would be a  valid  representation of the 26 coadded RCs. Each sphere tags the best-fit model of  a coadded RC. We realize a large  failure of the NFW  density profile. Nearly all of the coadded RCs lie far outside of the goal volume. 

We now discuss  the other  two minor, but cosmologically important components of baryonic matter in disk systems. A bulge contribution to the circular velocity is  present in (only) disk galaxies with the highest stellar mass. In Appendix \ref{Appendix: b} we include this component in the circular velocity model of  the coadded RCs relative to these objects. We find that  the resulting best-fit model agrees well with that  obtained from Eq. (\ref{modv}). The other component is a gaseous HI disk   in the outermost parts of the least luminous objects. In Appendix  \ref{Appendix:c} we present the model  for the coadded RCs  of the latter.  We add an HI gaseous disk contribution   and compare the resulting best-fit model with  that of Eq. (\ref{modv}). We find a good agreement in this case as well. This shows that including  these two minor baryonic components does not change the results of this work.

It is important to compare the results of our work with a previous  study of {\it \textup{individual}} RCs \citep{Korsaga.19}.  It is worth noticing that the coadded RCs with respect to  the individual RCs show on average higher values of $c$ and a stronger correlation with the halo mass as likely expected by the coaddition process, which has eliminated a part of the random errors of the individual RCs. In Fig. \ref{cVopt} we plot our 26 $c$ versus $V_{opt}$ along with those obtained from  individual RCs in  \citet{Korsaga.19}. The agreement further reinforces the statement of the gross inability of NFW halos to reproduce the observed disk kinematics.
\begin{figure}[t!]
\begin{center}
\qquad
\includegraphics[scale=0.6]{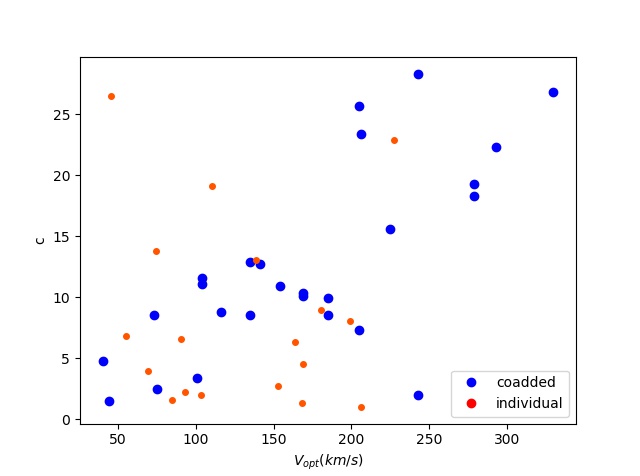}
\caption{Relation between concentration $c$ and optical velocity $V_{opt}$. The quantity $V_{opt}$ is derived using the relation
$M_B=-7 \log{V_{opt}-4.9}$ from \citet{Promareva.17}. The blue points are the results of this work, and the orange points are the results of \citet{Korsaga.19} using the fixed M/L fitting technique.}
\label{cVopt}
\end{center}
\end{figure}

%
\section{Conclusion}\label{conclu}
 More than 3200 RCs of disk systems with different $V_{opt}$, ranging from $20~km\ s^{-1}$ to $330~km\ s^{-1}$, were stacked into $26$ coadded RCs that extended out to $R_{opt}$ and beyond. These  data represent the whole kinematics of disk systems well and were used to investigate a particular mass distribution. This is characterized by a dark halo with an NFW profile and a stellar Freeman exponential thin disk. The model has three free parameters: the disk mass $M_D$, the halo virial mass $M_{vir}$ , and the halo concentration $c$ . These were derived by means of best-fitting the coadded RCs. Because the NFW velocity model is an analytical model that is valid from $r=0$ and $r=R_{vir}$,   we can derive all the related $\rho_s$,$r_s$, and $R_{vir}$ from the two halo parameters above.
 
 The aim of our investigation was to determine  whether dark halos around spirals possess  cuspy inner profiles. The outcomes crucially add to the results that were obtained in the past 20 years by means of the analysis of several dozen individual RCs. Although these have given an unequivocal answer against this possibility, they cannot be considered either entirely unbiased or complete.
 
 By means of the above  26 coadded RCs we have completed the investigation about the core-cusp question in disk systems of any morphology and luminosity (mass). We found that generally, the model under investigation  fails to reproduce the observational  data. Not only does it fit the coadded RCs pooly, implying high values for $\chi_{red}^2$, but in many cases, the best-fit  values of the structural parameters $c$, $M_{vir}$ and $M_D$ are also very different from the expectations from simulations and from measurements independent of RCs data. Noticeably, the model fails at  any reference velocity from ($V_{opt}=20~km\ s^{-1} $)  to  ($V_{opt}>300~km\ s^{-1} $). One exception may be for  objects with $190\ km\ s^{-1} \leq  V_{opt} \leq 210 \ km\ s^{-1} $ , for  which the model appears to  perform as well as the Burkert cored halo + Freeman disk model; in this case, only  individual HI  RCs with the optical velocity in this range and with very much  extended kinematics can solve this riddle \citep{Karukes.Sal.17}. We stress that according to our results,  we conclude that within the $\Lambda CDM$ scenario,  the corification of initial cusps  must occur in the smallest and largest disk systems.

 A detailed investigation shows that the failure of the model has different aspects: the concentration $c$ takes a wide range of values from $1.5$ to $28.3$ that is well outside the range found in simulations. The dark matter halo masses in the 50\% of the cases reach implausibly high values. In some cases, the values of the disk masses are ridiculously low. The $\chi^2_{red}$ value, on the other hand, can be acceptable for only 38\% of the cases. 
 
 From the best-fit parameters we reconstructed the model scaling relations $c- M_{vir}$ and $M_D-M_{vir}$ that emerge to clearly contradict those obtained from observations and from N-body simulations \citep{Salucci.Lap.07, Shankar.Lap.06, Bullock.Kol.01}. The 3D relation of $c$- $M_D/M_{vir}$- $\chi_{red}^2$ in Fig. \ref{3dplt} clearly shows that most of the best-fit values  lie outside the volume that indicates a successful representation of the kinematics of disk galaxies. 

This result must be gauged by the fact that,  in previous works, we found that the cored halo+ Freeman disk is able to reproduce all these 26 coadded RCs with very low $\chi^2_{red}$ values ($\simeq 1$) and  yields meaningful relations among the model structural parameters \citep[see][]{Karukes.Sal.17, Lapi.Sal.18, Dipaolo.Sal.18}.

  In this work, however,  we make no claim  about {\bf i)} the $\Lambda$CDM scenario, in that cores in the DM density might be produced during the cosmological evolution of galaxies, {\bf ii)} the DM distribution at large radii $r>100 \ kpc$, in galaxies of other Hubble types, in spirals at high redshifts $z>0.5, $ and in disk systems just after their formation for the evident lack of available measurements. Only a large number of high-quality individual RCs, that is, 100 per magnitude, per Hubble type, per 0.5 redshift can resolve all these questions.
\begin{acknowledgements}
We thank the referee for suggestions that have improved the quality of the paper 
and Chiara Di Paolo for useful discussions.
\end{acknowledgements}
\bibliographystyle{aa}
\bibliography{references}
\section*{SUPPORTING INFORMATION}
Supplementary data are available at A\&A online as Table.txt.

\begin{appendix}

\section{Best-fit rotation curves and triangle plots}\label{Appendix: a}

In this section we plot the 25 velocity best-fitting models to their corresponding coadded RCs (left panels of Fig. \ref{trg22}). The triangular plots for the 25 mass models are also shown in this figure (right panels of Fig. \ref{trg22}).  
\begin{figure*}
\begin{center}
\vspace*{2cm}
\subfloat[PSS96, bin 1]{\includegraphics[scale=.5]{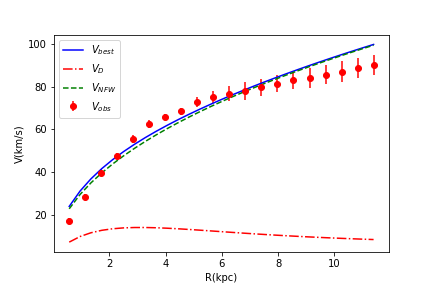}\hfill
\qquad
\includegraphics[scale=.5]{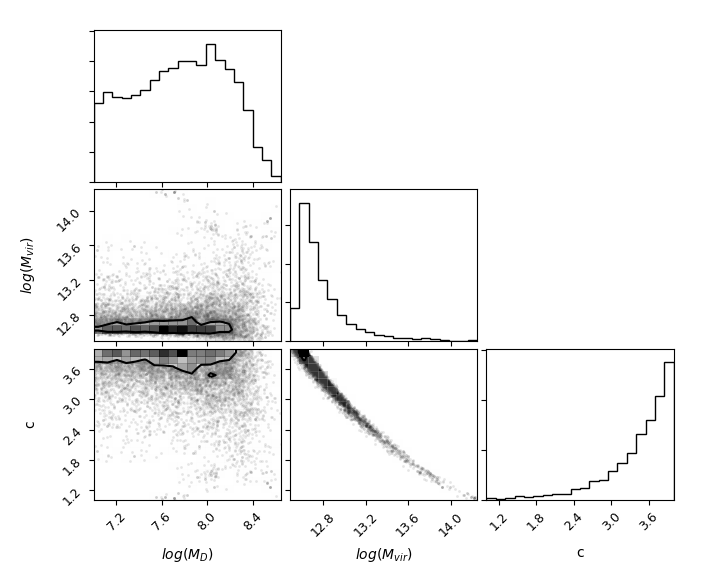}}\hfill
\vspace{2cm}
\subfloat[PSS96, bin 2]{\includegraphics[scale=.5]{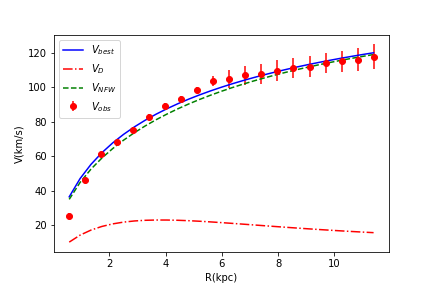}\hfill
\qquad
\includegraphics[scale=.5]{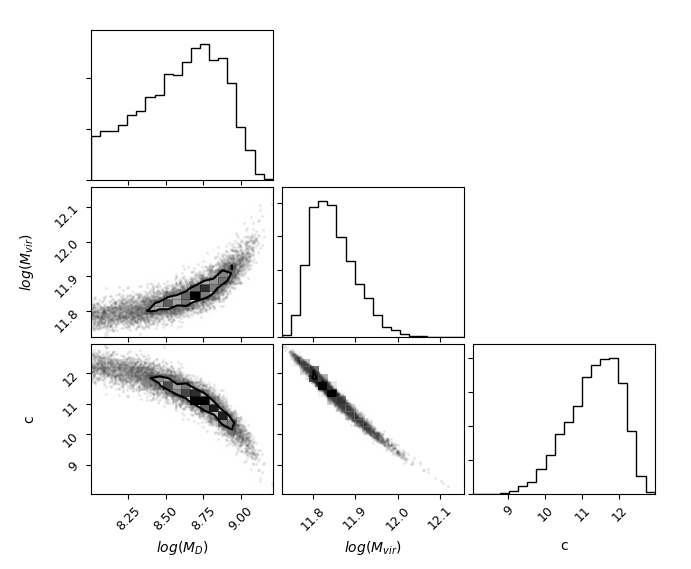}}\vfill
\vspace*{1cm}
\caption{\textit{continued}}
\end{center}
\end{figure*}

\begin{figure*}
\begin{center}
\ContinuedFloat
\vspace*{2cm}
\subfloat[PSS96, bin 3]{\includegraphics[scale=.5]{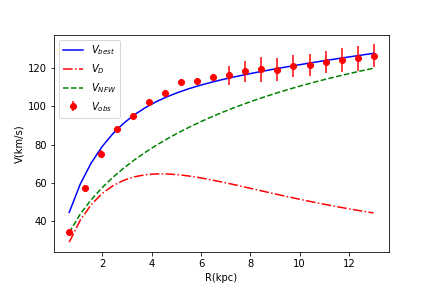}\hfill
\qquad
\includegraphics[scale=.5]{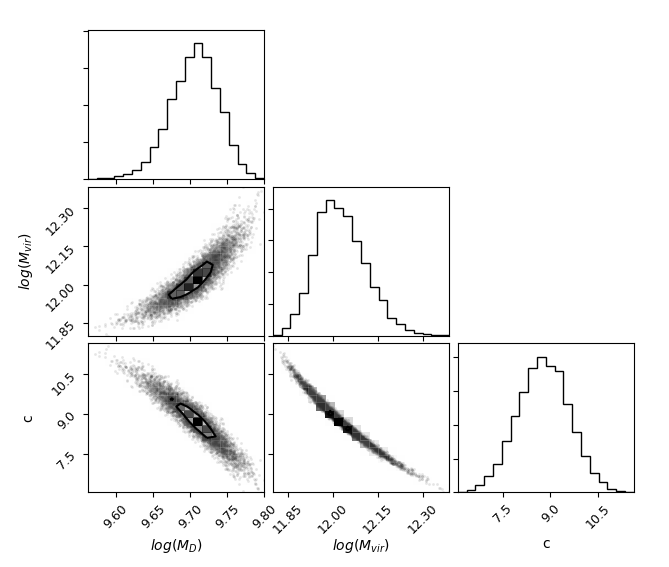}}\hfill
\vspace{2cm}
\subfloat[PSS96, bin 4]{\includegraphics[scale=.5]{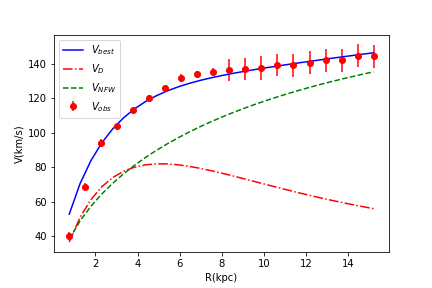}\hfill
\includegraphics[scale=.5]{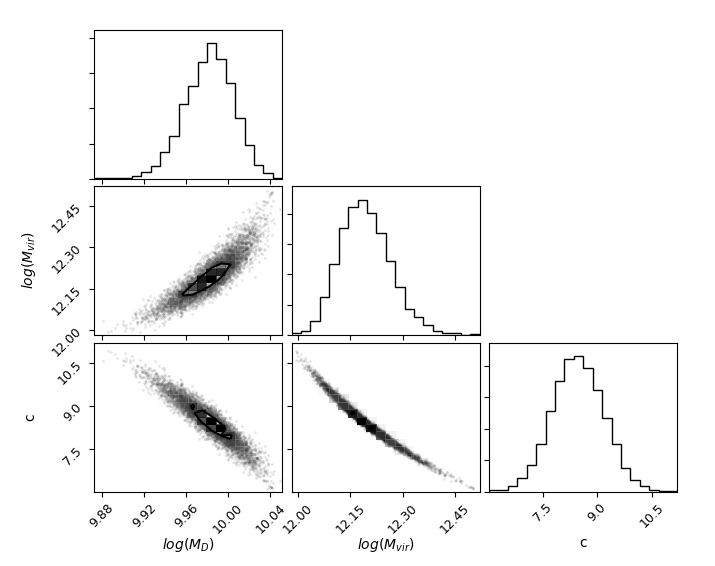}}\vfill
\vspace*{1cm}
\caption{\textit{continued}}
\end{center}
\end{figure*}

\begin{figure*}
\begin{center}
\ContinuedFloat
\vspace*{2cm}
\subfloat[PSS96, bin 5]{\includegraphics[scale=.5]{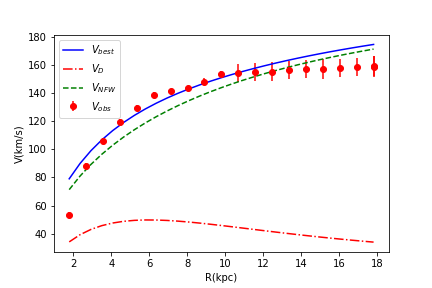}\hfill
\qquad
\includegraphics[scale=.5]{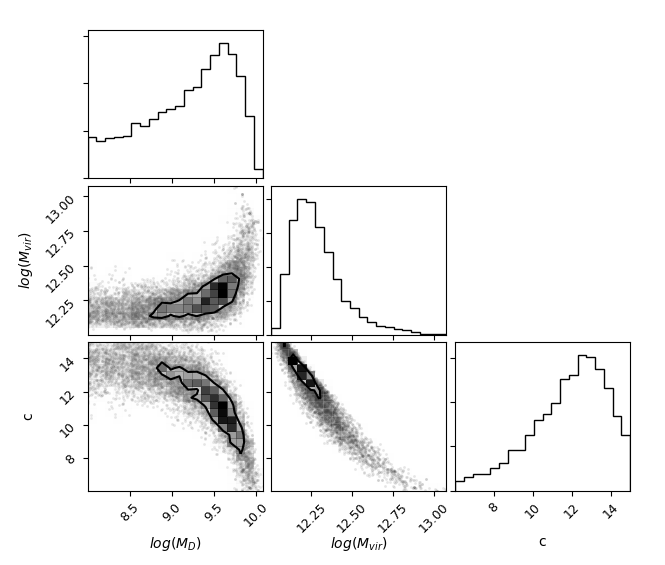}}\hfill
\vspace{2cm}
\subfloat[PSS96, bin 6]{\includegraphics[scale=.5]{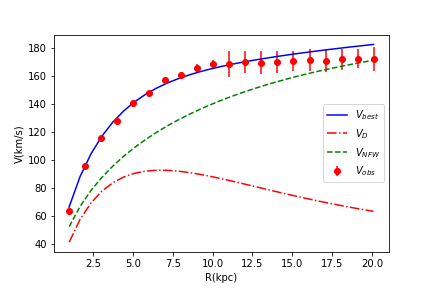}\hfill
\qquad
\includegraphics[scale=.5]{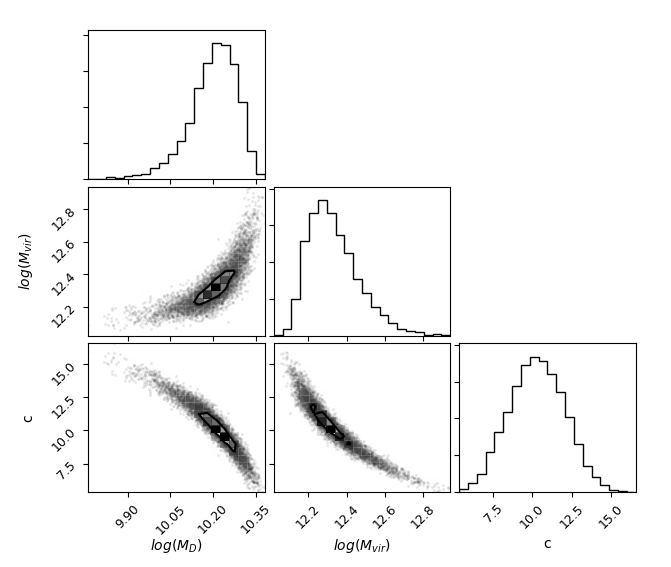}}\vfill
\vspace*{1cm}
\caption{\textit{continued}}
\end{center}
\end{figure*}

\begin{figure*}[!tp]
\begin{center}
\ContinuedFloat
\vspace*{2cm}
\subfloat[PSS96, bin 7]{\includegraphics[scale=.5]{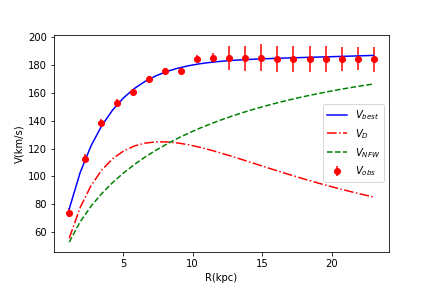}\hfill
\qquad
\includegraphics[scale=.5]{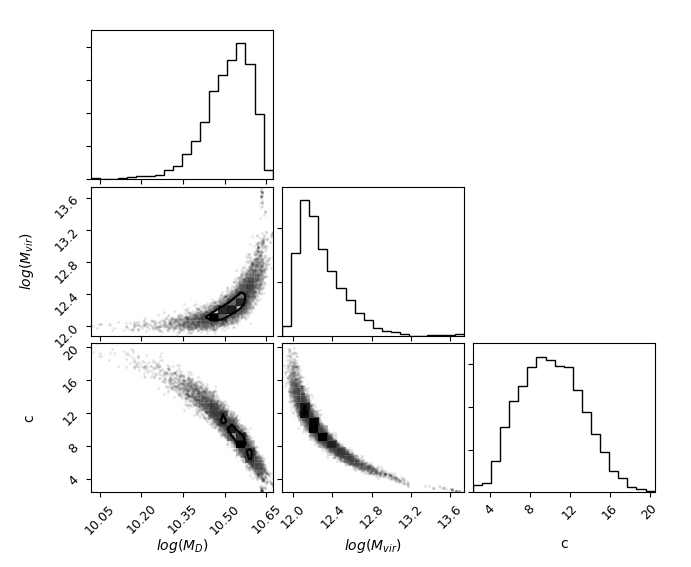}}\hfill
\vspace{2cm}
\subfloat[PSS96, bin 8]{\includegraphics[scale=.5]{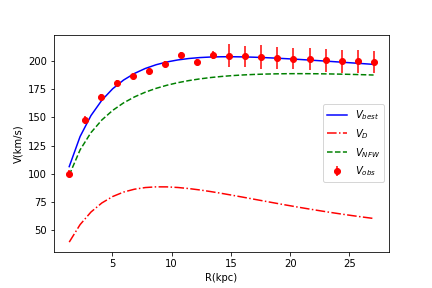}\hfill
\qquad
\includegraphics[scale=.5]{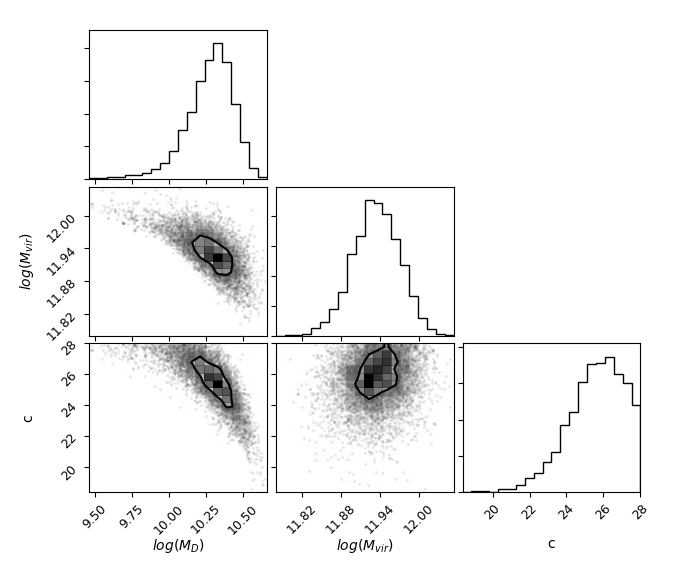}}\vfill
\vspace*{1cm}
\caption{\textit{continued}}
\end{center}
\end{figure*}

\begin{figure*}[!tp]
\begin{center}
\ContinuedFloat
\vspace*{2cm}
\subfloat[PSS96, bin 9]{\includegraphics[scale=.5]{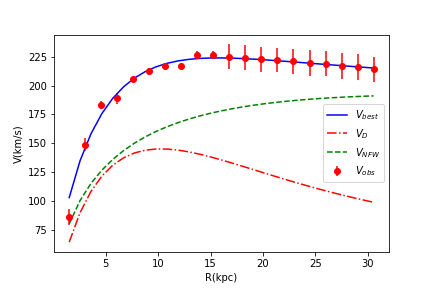}\hfill
\qquad
\includegraphics[scale=.5]{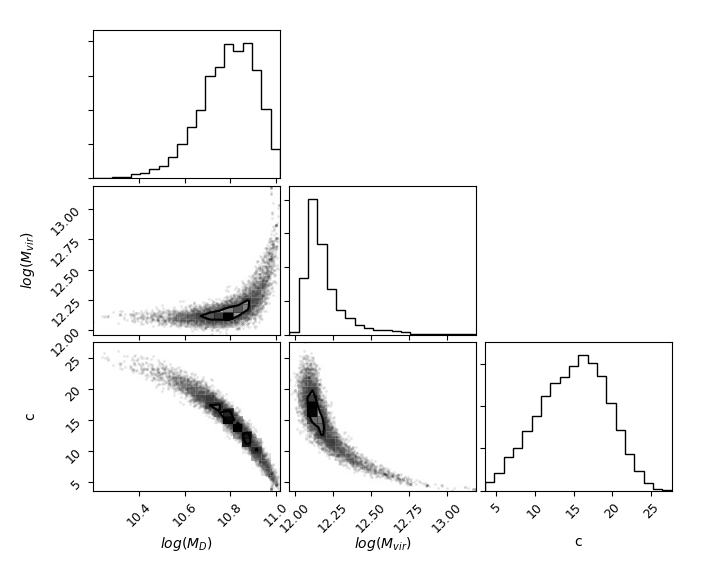}}\hfill
\vspace{2cm}
\subfloat[PSS96, bin 10]{\includegraphics[scale=.5]{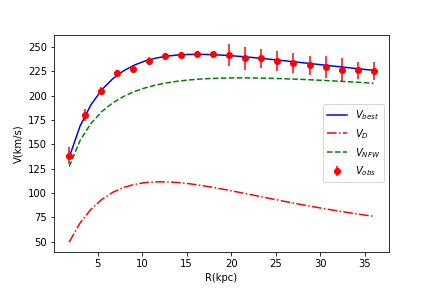}\hfill
\includegraphics[scale=.5]{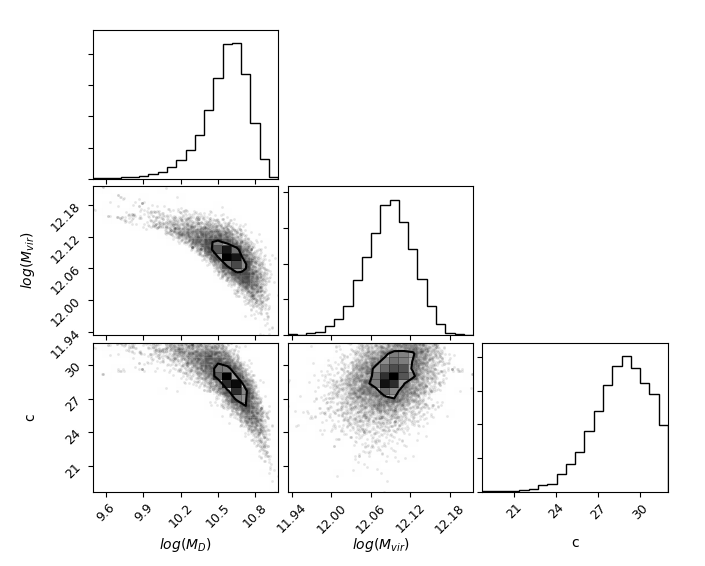}}\vfill
\vspace*{1cm}
\caption{\textit{continued}}
\end{center}
\end{figure*}

\begin{figure*}[!tp]
\begin{center}
\ContinuedFloat
\vspace*{2cm}
\subfloat[PSS96, bin 11]{\includegraphics[scale=.5]{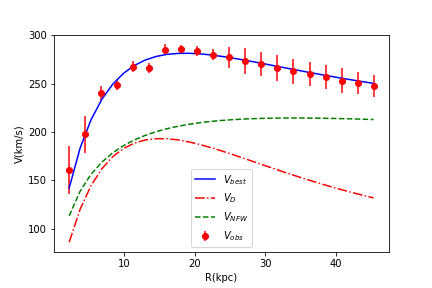}\hfill
\includegraphics[scale=.5]{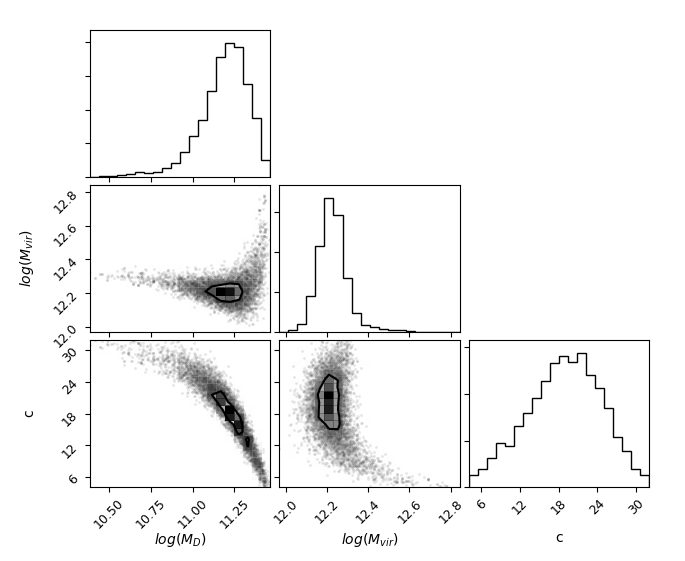}}
\vspace{2cm}
\subfloat[LSB, bin 1]{\includegraphics[scale=.5]{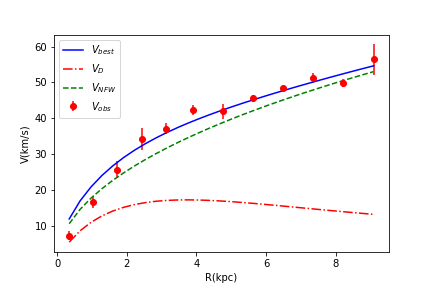}\hfill
\includegraphics[scale=.5]{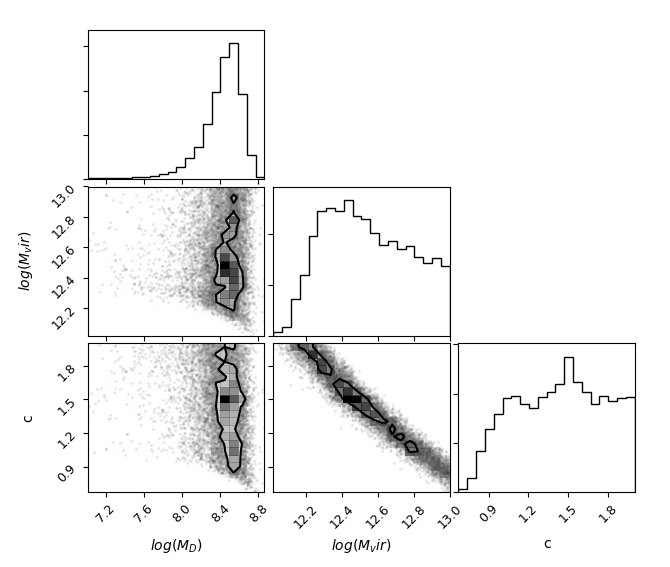}}\vfill
\vspace*{1cm}
\caption{\textit{continued}}
\end{center}
\end{figure*}

\begin{figure*}[!tp]
\begin{center}
\ContinuedFloat
\vspace*{2cm}
\subfloat[LSB, bin2]{\includegraphics[scale=.48]{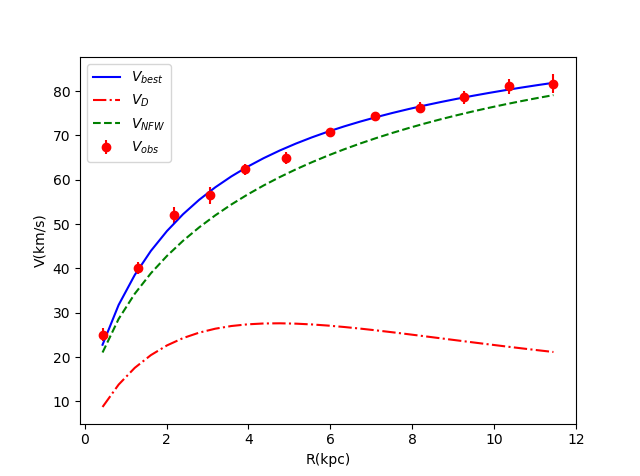}\hfill
\includegraphics[scale=.5]{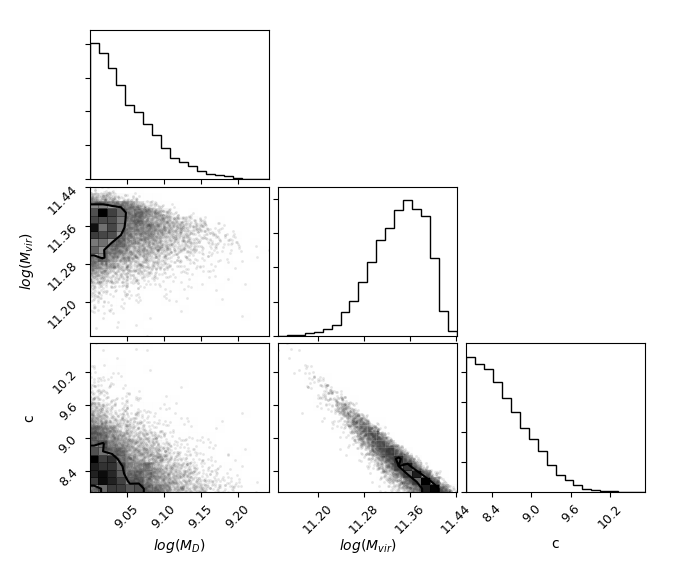}}
\vspace{2cm}
\subfloat[LSB, bin 3]{\includegraphics[scale=.53]{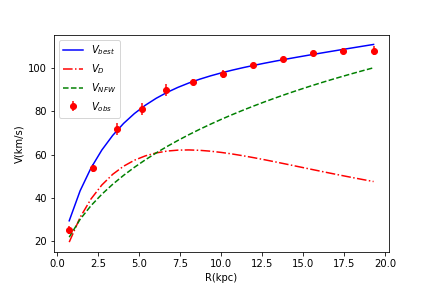}\hfill
\includegraphics[scale=.5]{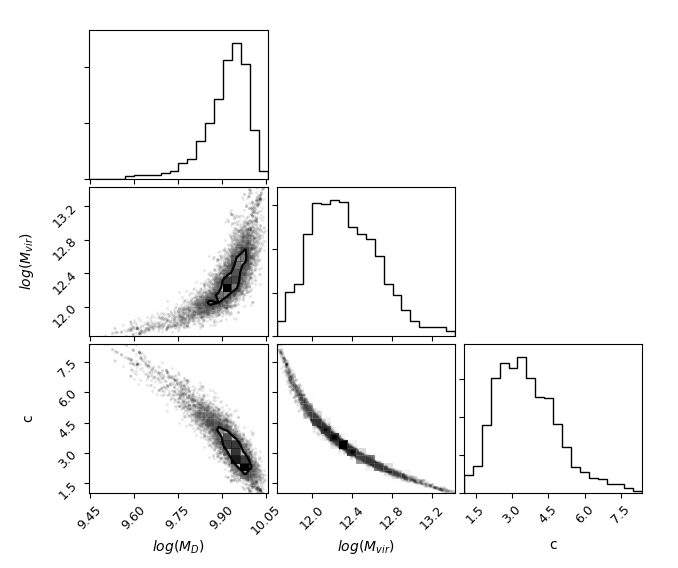}}\vfill
\vspace*{1cm}
\caption{\textit{continued}}
\end{center}
\end{figure*}

\begin{figure*}[!tp]
\begin{center}
\ContinuedFloat
\vspace*{2cm}
\subfloat[LSB, bin 4]{\includegraphics[scale=.5]{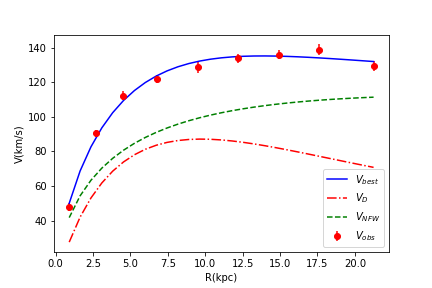}
\hfill
\includegraphics[scale=.5]{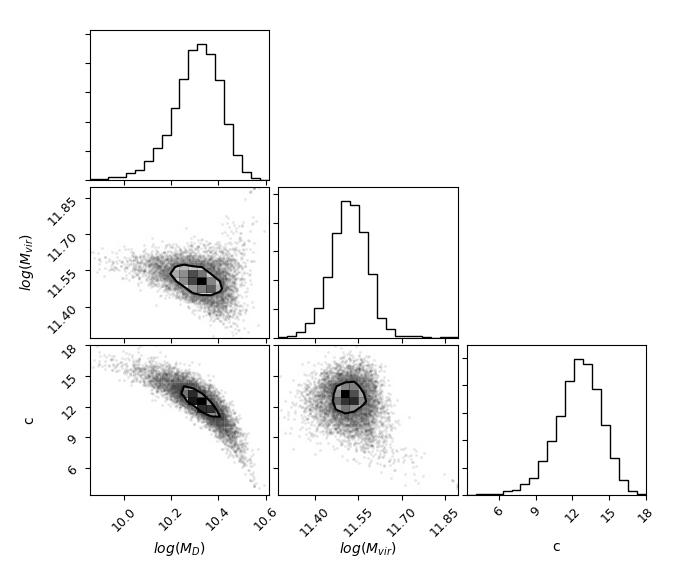}}\hfill
\vspace{2cm}
\subfloat[LSB, bin 5]{\includegraphics[scale=.5]{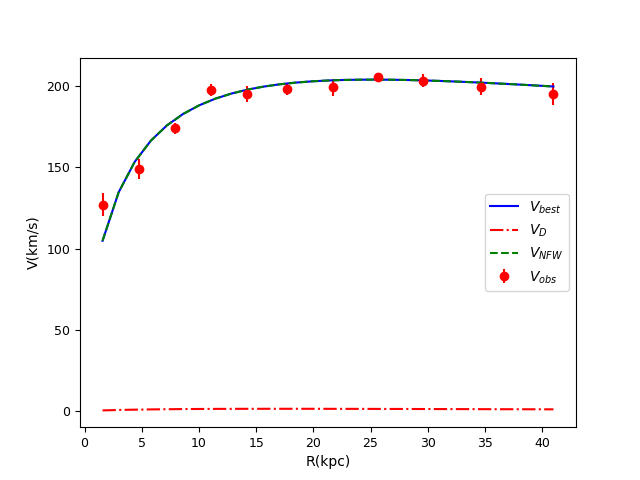}
\hfill
\includegraphics[scale=.5]{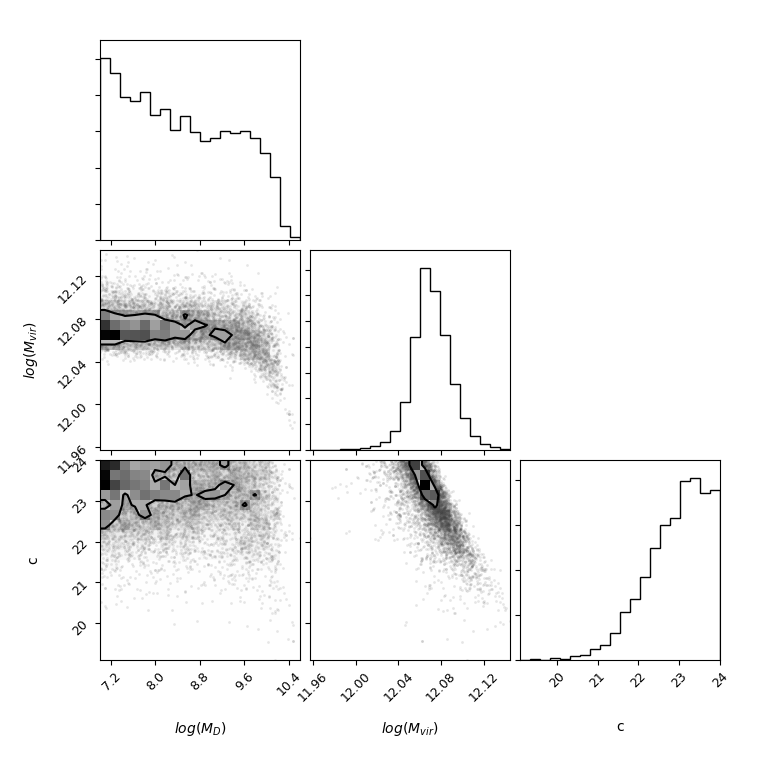}}\vfill
\vspace*{1cm}
\caption{\textit{continued}}
\end{center}
\end{figure*}

\begin{figure*}[!tp]
\begin{center}
\ContinuedFloat
\vspace*{2cm}
\subfloat[Ca06, bin 1]{\includegraphics[scale=.5]{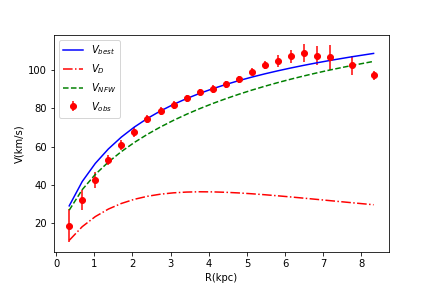}\hfill
\includegraphics[scale=.5]{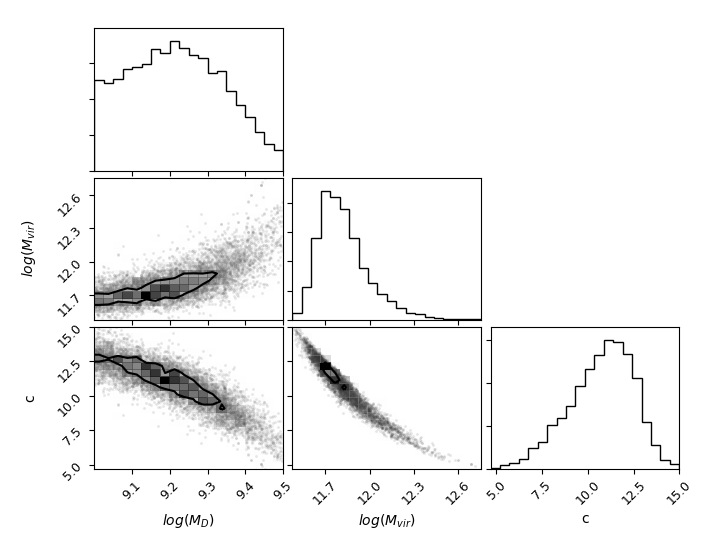}}
\vspace{2cm}
\subfloat[Ca06, bin 2]{\includegraphics[scale=.5]{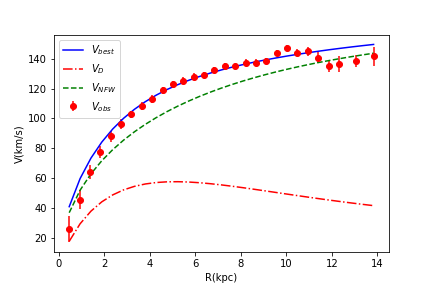}\hfill
\includegraphics[scale=.5]{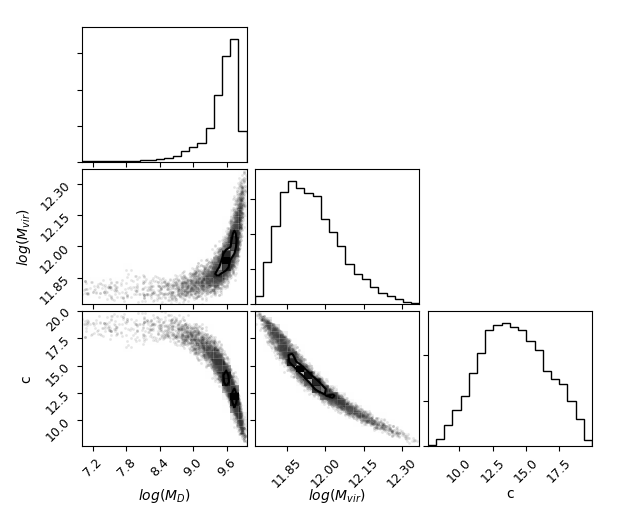}}\vfill
\vspace*{1cm}
\caption{\textit{continued}}
\end{center}
\end{figure*}

\begin{figure*}
\begin{center}
\ContinuedFloat
\vspace*{2.2cm}
\subfloat[Ca06, bin 3]{\includegraphics[scale=.5]{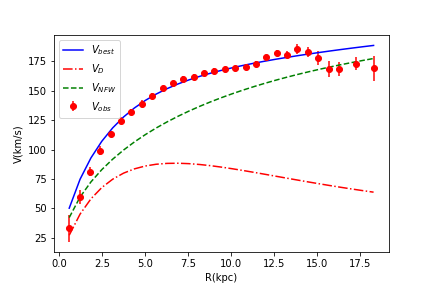}\hfill
\includegraphics[scale=.5]{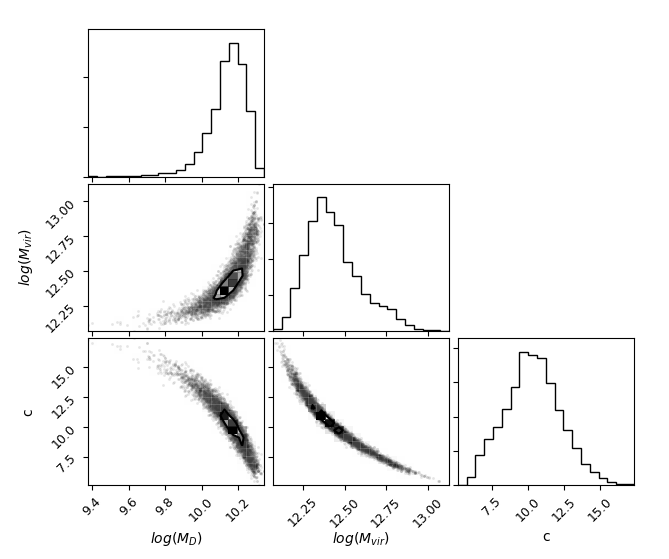}}\hfill
\vspace*{1cm}
\subfloat[Ca06, bin 4]{\includegraphics[scale=.5]{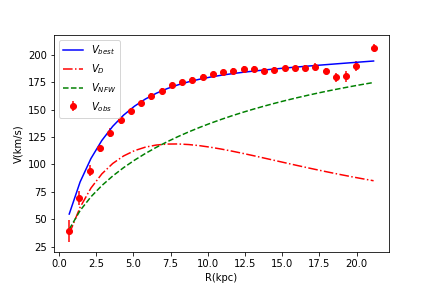}\hfill
\includegraphics[scale=.5]{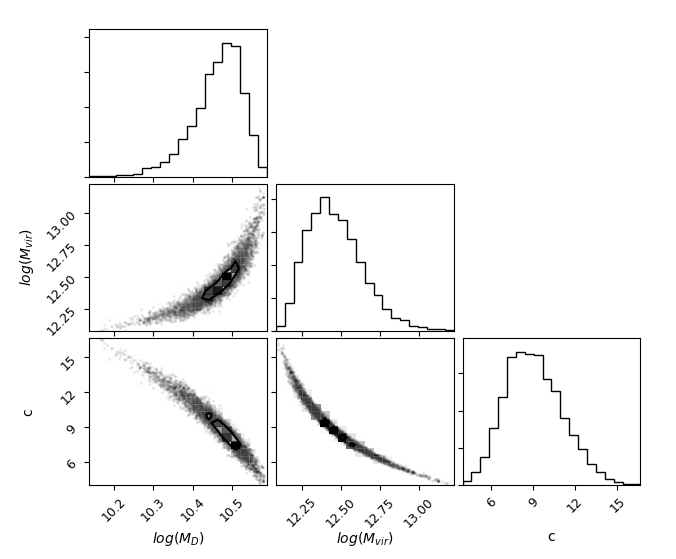}}\vfill
\vspace*{1cm}
\caption{\textit{continued}}
\end{center}
\end{figure*}

\begin{figure*}[!tp]
\begin{center}
\ContinuedFloat
\vspace*{2cm}
\subfloat[Ca06, bin 5]{\includegraphics[scale=.5]{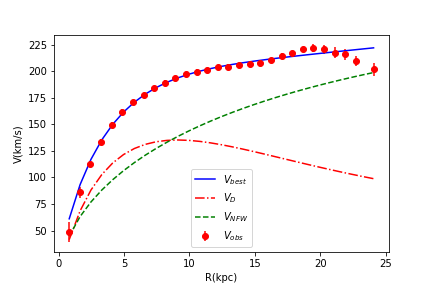}\hfill
\includegraphics[scale=.5]{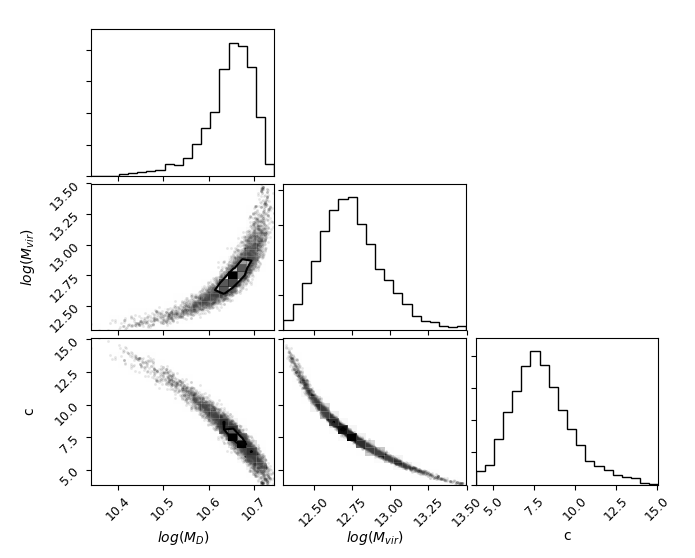}}\hfill
\vspace*{1cm}
\subfloat[Ca06, bin 6]{\includegraphics[scale=.5]{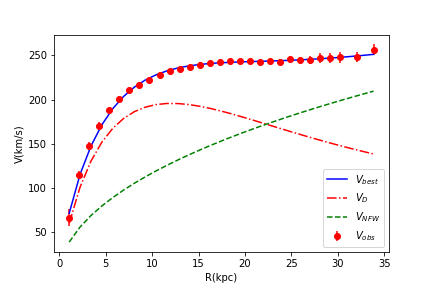}\hfill
\includegraphics[scale=.5]{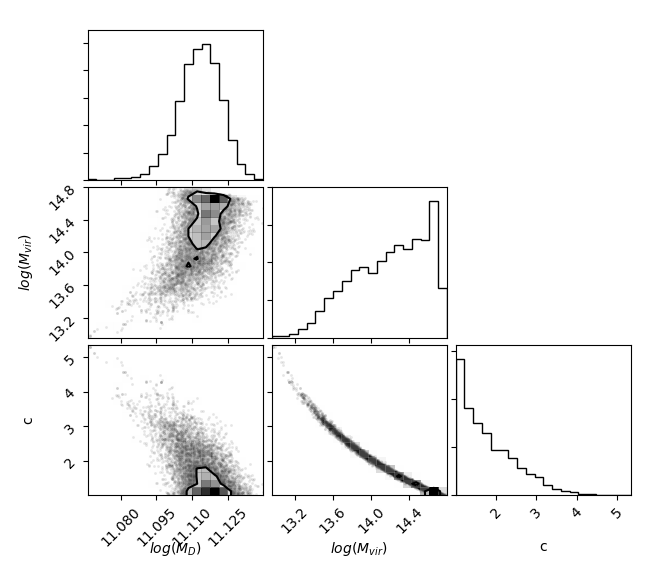}}\vfill
\vspace*{1cm}
\caption{\textit{continued}}
\end{center}
\end{figure*}

\begin{figure*}[!tp]
\begin{center}
\ContinuedFloat
\vspace*{2cm}
\subfloat[Ca06, bin 7]{\includegraphics[scale=.5]{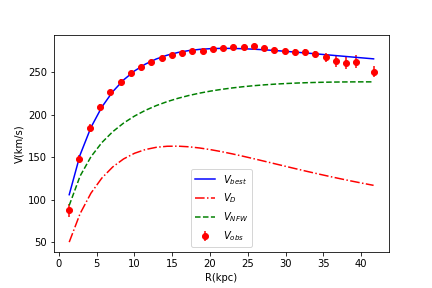}\hfill
\includegraphics[scale=.5]{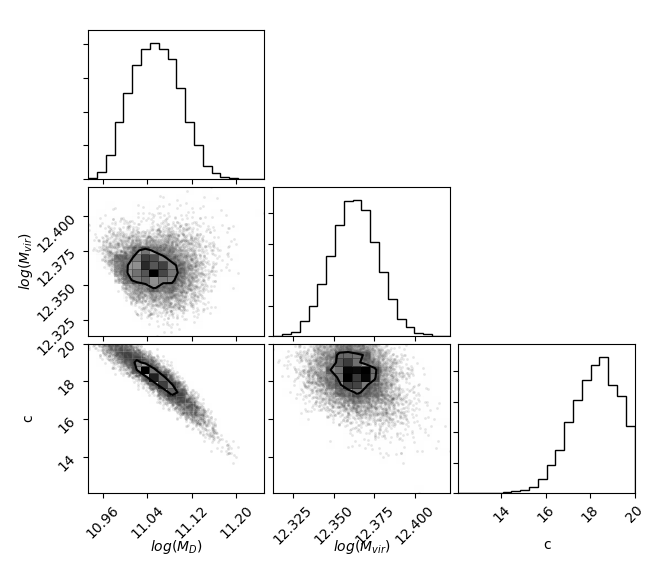}}\vfill
\vspace*{1cm}
\subfloat[Ca06, bin8]{\includegraphics[scale=.5]{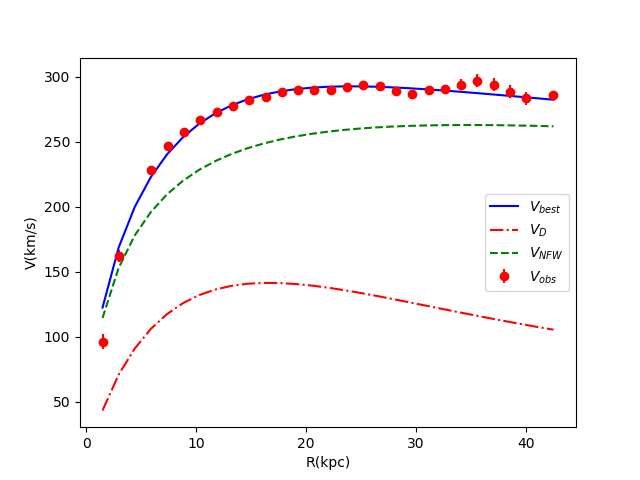}\hfill
\includegraphics[scale=.5]{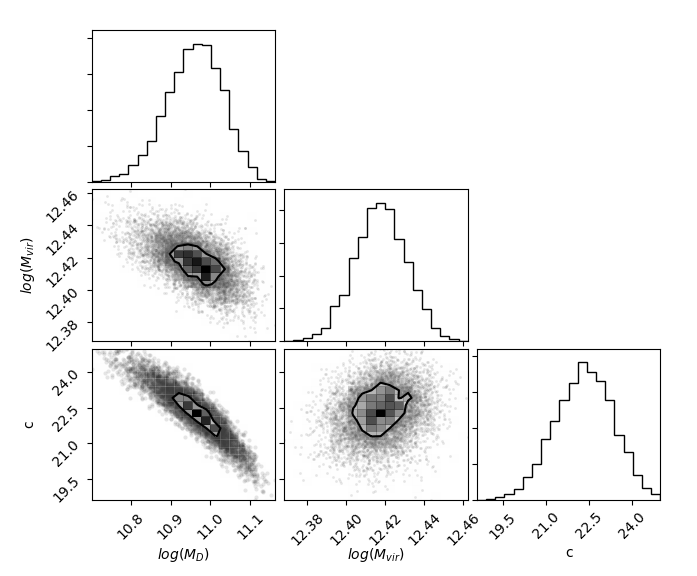}}\vfill
\vspace*{1cm}
\caption{\textit{continued}}
\end{center}
\end{figure*}

\begin{figure*}[h!]
\begin{center}
\ContinuedFloat
\vspace*{2cm}
\subfloat[Ca06, bin9]{\includegraphics[scale=.5]{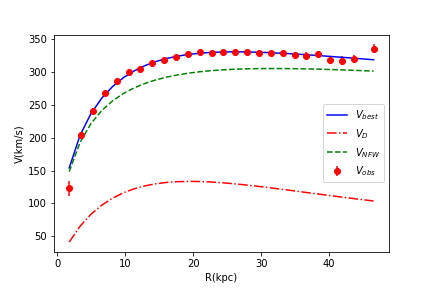}
\hfill
\includegraphics[scale=.5]{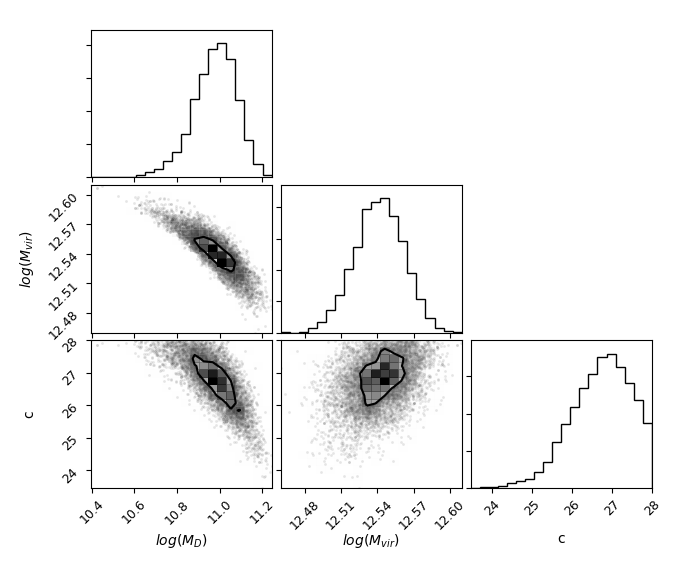}}
\vfill
\caption{RC modeling based on the NFW halo profile and the stellar Freeman exponential disk (left panels) and their corresponding triangle plots of 2D invariance between parameters with 1$\sigma$ uncertainty (right panels). In the left panel the red circles are observed data with error bars, the blue line is the best-fitting model, the dashed green line shows the NFW halo component, and the dot-dashed red line is the disk component.}\label{trg22}
\vfill
\end{center}
\end{figure*}

\section{Bulge contribution to the rotation curves of disk systems}
\label{Appendix: b}

The largest disk galaxies with $V_{opt} > 220\ km\ s^{-1}$ have an additional inner stellar component, a bulge.  The rotation curve of these systems  is then composed  of a spherical bulge, a disk, and a dark halo,
\begin{align*}
V^2(R)=V_B^2(R)+V_D^2(R)+V_{NFW}^2(R),
\end{align*}
where $V_D$ and $V_{NFW}$ are the disk and dark halo components of the rotation curve, respectively, and are defined in Eqs. (\ref{vd0}) and (\ref{VNFW}). $V_{B}$ is the bulge contribution. This component is smaller than the stellar disk  by at least a factor of 3.  We assumed that we can represent this component by the following velocity profile: 
\begin{align*}
V^2_B(R)= \frac{GM_{B}}{R}\frac{x^2}{\left(0.1^2+x^2\right)},
\end{align*}

where $M_{B}$ is the bulge mass. In the above, the bulge behaves like a point mass at large distances, for $R<<1/3\, R_D$ , it increases like $ R^2$.

The velocity model therefore  has four free parameters.  Table \ref{appendix:apenb} shows the best-fitting results for the most massive bins of the three samples \textbf{Ca06}, \textbf{PSS96,} and \textbf{LSB} where  this component is (mildly) relevant. The corresponding RCs are shown in Fig. \ref{RCb1}. The bold and red values are the same as in Tables (\ref{rePSS}-\ref{reCa}). We realize that including the very central bulge does not change the results of this work. It is worth noting that we did not investigate the
bulge-dominated Sa spirals here.

\begin{table*}[!h]
\begin{center}
\caption{Values of the best-fit parameters of the mass model along with  their $1\sigma$ uncertainties. Column $(1)  $ lists the sample name, Col. $(2)$ the optical radius, Col. $(3)$ the optical velocity, Col. $(4)$ the concentration, Col. $(5)$ the bulge mass, Col. $(6)$ the disk mass, Col. $(7)$ the halo mass (virial mass), and Col. $(8)$ the reduced $\chi^2$ value.}\label{appendix:apenb}
\begin{tabular}{cccccccc}
\hline
 $sample$ & $R_{opt}$ & $V_{opt}$ & $c$ & $\log (M_B)$ & $\log (M_D)$ & $\log(M_{vir})$ & ${\chi^2_{red}}$ \\[0.8ex]
 $-$ & $(kpc)$ & $(km\ s^{-1}) $ & $-$ & $(M_\odot)$ & $(M_\odot) $ & $(M_\odot)$ & $-$\\[0.8ex]
 $\textbf{(1)}$ & $\textbf{(2)}$ & $\textbf{(3)}$ & $\textbf{(4)}$ & $\textbf{(5)}$ & $\textbf{(6)}$& $\textbf{(7)}$ & $\textbf{(8)}$ \\ \hline
 $PSS96$ & $22.7$ & $279$ & $\textbf{12.52}^{+5.2}_{-5.6}$ & $\textcolor{red}{9.84^{+\infty}_{-1.1}}$ & $11.29^{+0.08}_{-0.1}$ & $12.28^{+0.2}_{-0.09}$ & $0.47$ \\ \hline
 $Ca06$ & $29.1$ & $330$ & $\textbf{1.8}^{+1.5}_{-0.05}$ & $10.46^{+0.05}_{-0.04}$ & $11.63^{+0.01}_{-0.01}$ & $\textbf{14.06}^{+\infty}_{-0.01}$ & $\textbf{2}$ \\ \hline
 $LSB$ & $25.3$ & $206$ & $6.7^{+5}_{-5}$ & $10.13^{+0.1}_{-0.3}$ & $10.9^{+0.2}_{-0.6}$ & $12.24^{+0.7}_{-0.2}$ &  $\textbf{1.7}$ \\ 
 \hline
\end{tabular}
\end{center}
\end{table*}

\begin{figure*}
\begin{center}
\vspace*{2cm}
\subfloat[PSS96, bin 11]{\includegraphics[scale=.5]{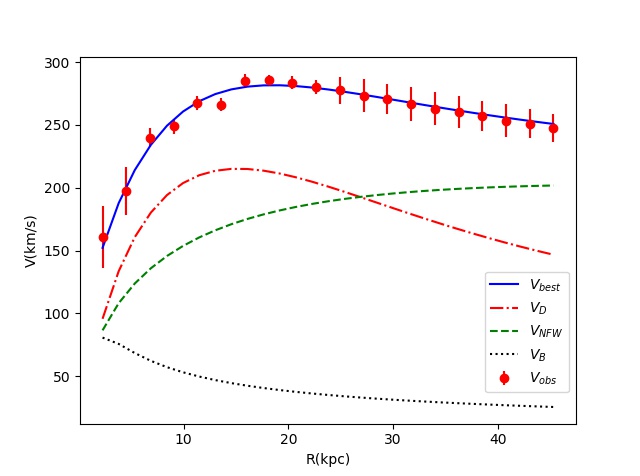}\hfill
\qquad
\includegraphics[scale=.45]{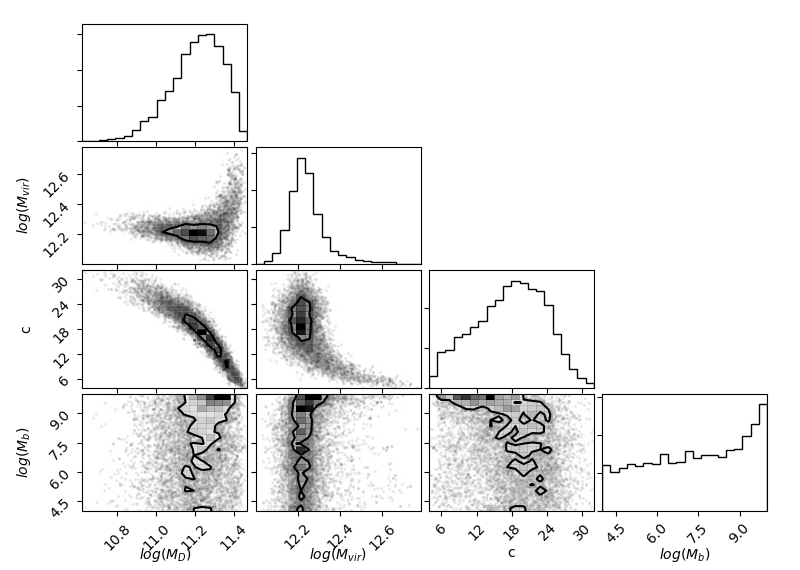}}\hfill
\vspace{2cm}
\subfloat[LSB, bin 5]{\includegraphics[scale=.5]{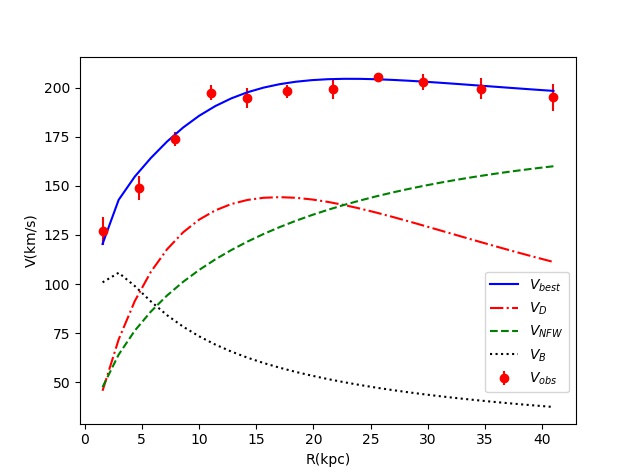}\hfill
\qquad
\includegraphics[scale=.5]{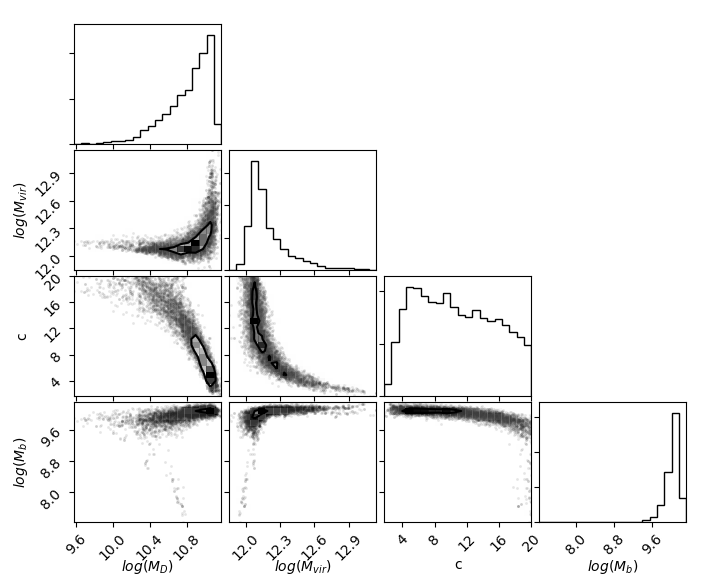}}\vfill
\vspace*{1cm}
\caption{\textit{continued}}
\end{center}
\end{figure*}

\begin{figure*}
\begin{center}
\ContinuedFloat
\vspace*{2cm}
\subfloat[Ca06, bin 9]{\includegraphics[scale=.5]{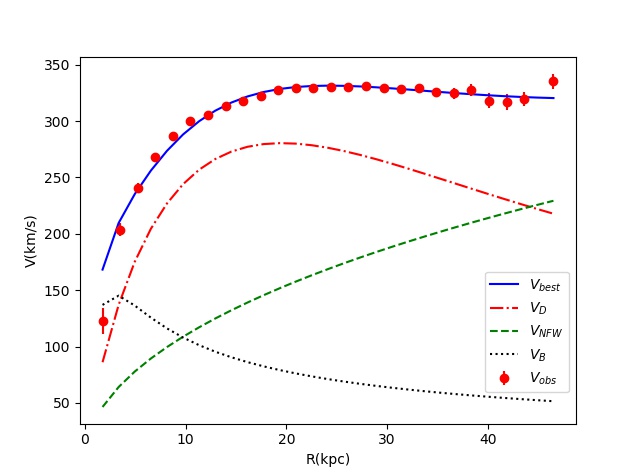}\hfill
\qquad
\includegraphics[scale=.45]{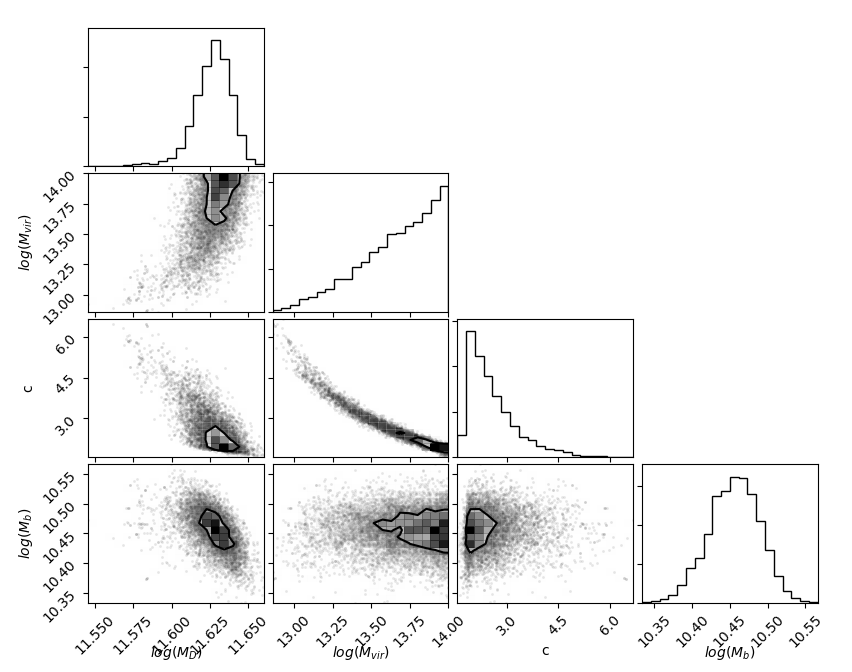}}
\vspace{1cm}
\caption{Observed and fit RC modeling: The components of the NFW dark halo (green dashed), disk (dot-dashed red), and bulge (dashed black) that contribute to the total rotation curve and the observed data (red circles) with error bars are shown in the left panels. Their corresponding triangle plots of the 2D invariance between parameters with 1$\sigma$ uncertainty are shown in the right panels.}\label{RCb1}
\end{center}
\end{figure*}

\section{Gas contribution to the rotation curves}\label{Appendix:c}

Disk galaxies also receive an HI contribution to their circular velocities. We did not consider this here because \textbf{i)} this component is relevant only in small objects and outside $R_{opt}$,  \textbf{ii)} in any case, the effects of incorporating this gaseous component results in NFW cusps that are even more unlikely \citep[e.g.,][]{Dipaolo.Sal.18, Karukes.Sal.17}.

\begin{figure*}
\begin{center}
\includegraphics[scale=0.6]{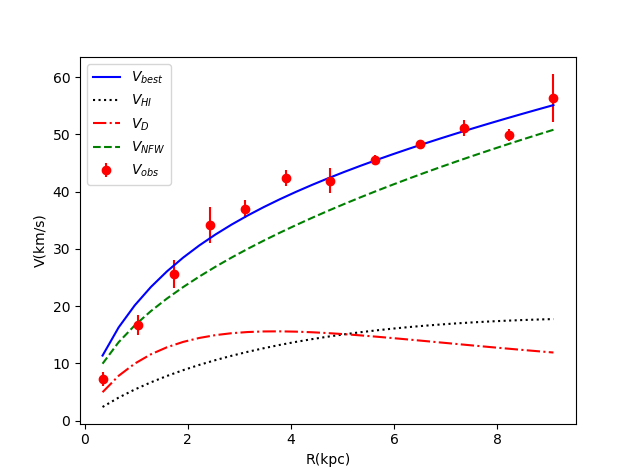}\hfill
\caption{Best-fit RC of the first bin of the   \textbf{LSB} sample by considering the gaseous component. The dashed green, dot-dashed red, dotted black, and solid blue lines stand
for the DM halo, the stellar disk, the gaseous disk, and the total contributions to the circular velocity.}
\label{gaslsb}
\end{center}
\end{figure*}

In this section we add these component to the stellar disk + NFW halo  model to the circular velocity of galaxies for the lowest velocity bins of our sample. The HI surface density can approximately be reproduced by a Freeman distribution with a scale length $R_{HI}$
that is about three times larger than that of the stellar disk \citep[see][]{Evoli.Sal.11,Wang.Fu.14}.  
 We then take for the contribution of the HI disk to the circular velocity
\begin{align*}
V_{HI}^2(R)=\frac{G M_{HI}}{6 R_D}\left[1.07\  x\right]^2\left[I_0 K_0- I_1 K_1\right],
\end{align*}
where $x=\frac{R}{R_{opt}}$ and $M_{HI}$ is the mass of HI disk, considered as a free parameter as $M_D$ and $I_n$ and $K_n$ are the modified Bessel functions calculated at $0.53 x$. The expected range of the free parameter values of the stellar+ gaseous disk  and the dark halo are reported in Table \ref{HIres} , and the mass model best-fitting result is shown in Fig. \ref{gaslsb}. Compared to the results we obtained when we  neglected the  gas contribution (see Fig. \ref{trg22}), it produces only a mild difference in the mass model when we include the gas.

\begin{table*}[!h]
\begin{center}
\caption{Coadded RC model result with gaseous disk. Masses are expressed in solar mass units.}
\begin{tabular}{cc}
\hline
 free parameter & range of values \\[0.8ex]
\hline
$\log(M_D)$ & $8-9$ \\ \hline
$\log(M_{vir})$ & $12.5-13.5$ \\ \hline
$c$ & $0.5-2$ \\ \hline
$\log(M_{HI})$ & $8.4-9.4$ \\ \hline
\end{tabular}\label{HIres}
\end{center}
\end{table*}

\end{appendix}
\end{document}